\documentclass[aps,twocolumn,superscriptaddress,footinbib,nobalancelastpage,prl]{revtex4-2}
\usepackage{amsmath,amssymb,amsfonts}
\usepackage[colorlinks=true,citecolor=blue,linkcolor=red,urlcolor=red]{hyperref}
\usepackage{graphicx}
\usepackage{braket}
\usepackage[normalem]{ulem}
\usepackage{xcolor}

\newcommand*{\s}[2]{S_{#2}^{#1}} 
\newcommand{\erfc}{\mathrm{erfc}}

\begin{document}

\title{Arresting classical many-body chaos by kinetic constraints}

\author{Aydin Deger}
\email{a.deger@lboro.ac.uk} 
\affiliation{Interdisciplinary Centre for Mathematical Modelling and Department of Mathematical Sciences,
  Loughborough University, Loughborough, Leicestershire LE11 3TU,
  UK}

\author{Sthitadhi Roy}
\email{sthitadhi.roy@icts.res.in} 
\affiliation{
  International Centre for Theoretical Sciences, Tata Institute of Fundamental Research, Bengaluru 560089, India
}

\affiliation{Rudolf Peierls Centre
  for Theoretical Physics, Oxford University,
  Parks Road, Oxford OX1 3PU, UK} 

  \affiliation{Physical
  and Theoretical Chemistry, Oxford University, South Parks Road,
  Oxford OX1 3QZ, UK} 

\author{Achilleas Lazarides}
\email{a.lazarides@lboro.ac.uk} 
\affiliation{Interdisciplinary Centre for Mathematical Modelling and Department of Mathematical Sciences,
  Loughborough University, Loughborough, Leicestershire LE11 3TU,
  UK}

\begin{abstract}
  We investigate the effect of kinetic constraints on classical many-body chaos in a translationally-invariant Heisenberg spin chain using a classical counterpart of the out-of-time-ordered correlator (OTOC). The strength of the constraint drives a \textit{dynamical phase transition} separating a delocalised phase, where the classical OTOC propagates ballistically, from a localised phase, where the OTOC does not propagate at all and the entire system freezes. This is unexpected given that all spin configurations are dynamically connected to each other. We show that localisation arises due to the dynamical formation of frozen islands, contiguous segments of spins immobile due to the constraints, dominating over the melting of such islands.
\end{abstract}

\maketitle

Ergodicity lies at the heart of bridging microscopic theories of many-body systems to macroscopic thermodynamic descriptions of such systems via their statistical mechanics~\cite{landaulifshitz}. Much of this is underpinned by the notion of chaos captured by the butterfly effect -- an infinitesimal local change in the initial condition, a wingbeat of the proverbial butterfly, can amplify exponentially and spread ballistically in spacetime to effect drastic changes in the global state at later times, such as cause a tornado in a different part of the world~\cite{lorenz1963deterministic,lorenz1995,lorenz2001butterfly}.

Recently, the out-of-time-ordered correlator (OTOC) has emerged as a prominent diagnostic for many-body quantum chaos~\cite{larkin1969quasisclassical,shenker2014black,roberts2015diagnosing,maldacena2016bound,swingle2016measuring,bohrdt2017scrambling,luitz2017information,nahum2018operator,keyserlingk2018operator,rakovszky2018diffusive,khemani2018velocity}. Defined as $\mathcal{O}(x,t)=-\braket{[W(x,t),V(0,0)]^2}$, it measures the effect on a local operator $W$ at position $x$ and time $t$ of perturbing the system with an operator $V$ at $x=0$ and $t=0$. A classical counterpart of the OTOC was developed to characterise spatiotemporal chaos in classical many-body systems~\cite{das2018light,bilitewski2018temperature,murugan2021many,liu2021butterfly,bilitewski2021classical}. It quantifies how the degrees of freedom in two copies of a system dynamically decorrelate in spacetime due to an infinitesimal local difference in their initial conditions. For chaotic systems, the OTOC or its classical counterpart has a ballistically propagating front which might be sharp or broaden diffusively.

\begin{figure}
  \includegraphics[width=\linewidth]{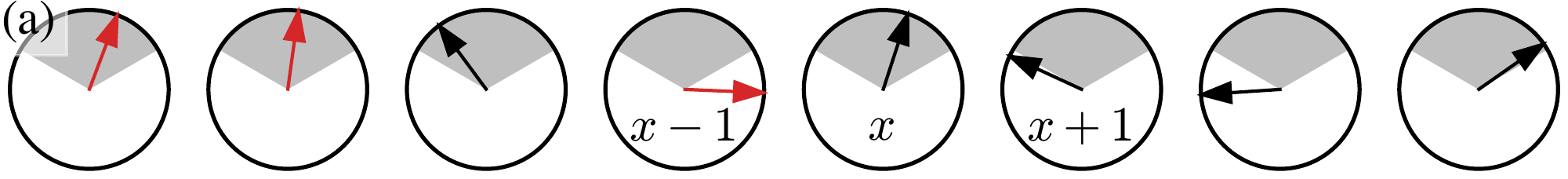}
  \includegraphics[width=\linewidth]{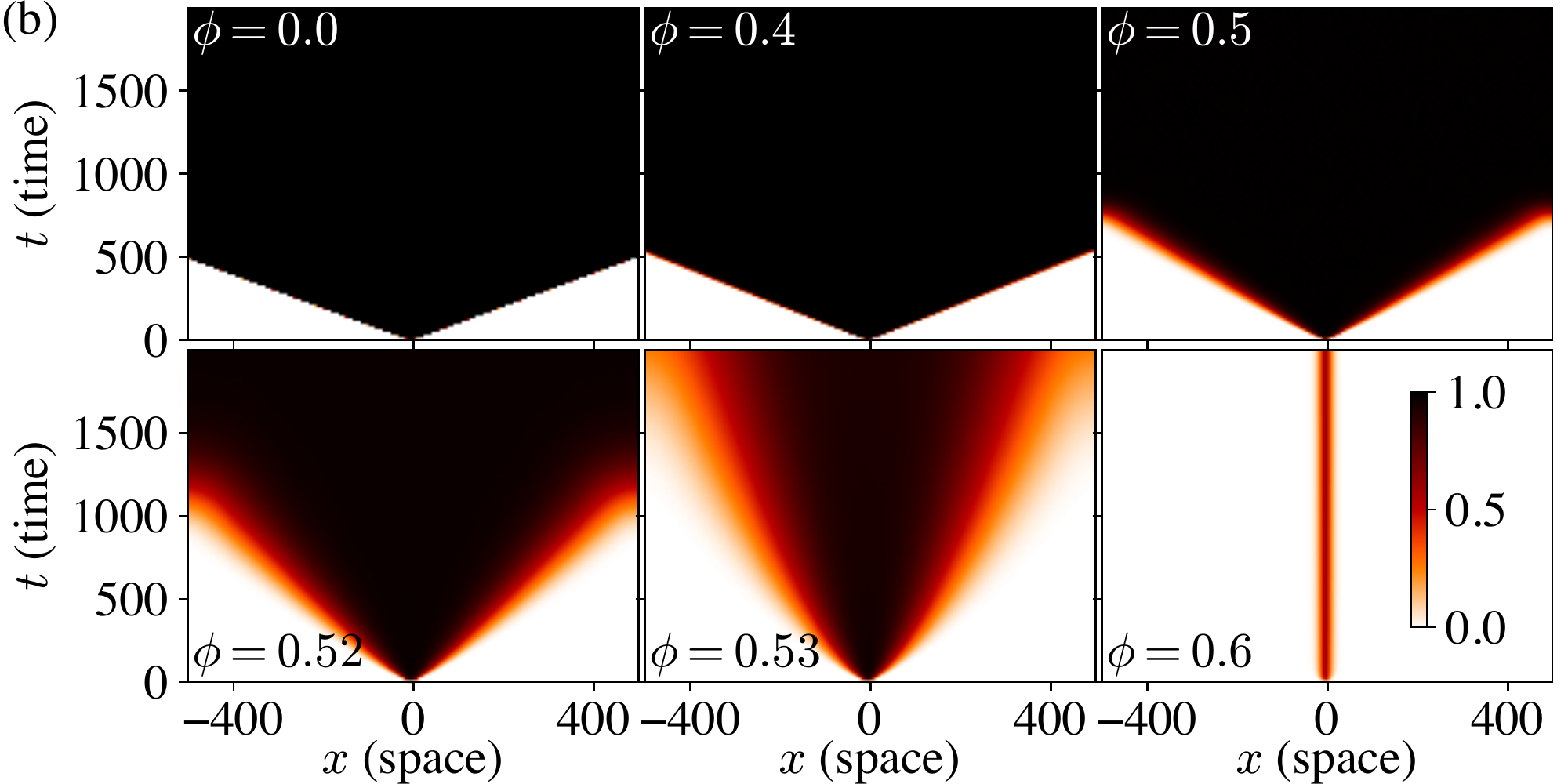}
  \caption{
  (a) Schematic of the constrained spin chain. The red spins are frozen as \emph{both} their neighbours lie inside the spherical sector (gray shade) which has a polar angle $\phi\pi$, whereas the black spins are free to evolve. (b) The classical OTOCs/decorrelators, $\braket{D(x,t)}$, as colourmaps in spacetime for several values of $\phi$. Note the rapid slowing down of the light-cone with $\phi$ near $\phi_c\approx 0.53$ above which the light-cone is fully arrested.}
  \label{fig:otocs}
  \end{figure}

Although ergodicity is generally considered the default, it is by now clear that there exist several classes of systems where ergodicity is broken, often strongly and robustly. Such systems therefore violate conventional statistical mechanics and thermodynamics; fundamental questions thus emerge about their universal, macroscopic descriptions. Telltale signatures of ergodicity breaking include long, often divergent, relaxation timescales, absence of thermalisation, suppressed transport, and specific to quantum systems, arrested growth of entanglement. All of these above phenomena can be loosely brought under the umbrella of (quasi)localisation of classical or quantum information.

Among the earliest examples of ergodicity breaking, perhaps the most significant ones are structural and spin glasses which become non-ergodic usually at low temperatures~\cite{gibbs1958nature,jackle1986models,  mezard2000statistical,edwards1975theory,sherrington1975solvable,binder1986spin,mezard1986spin}. A key ingredient in these systems is the presence of quenched random disorder. At the same time, disorder in low-dimensional quantum systems is also understood to cause strong ergodicity breaking, even at infinite temperatures, via Anderson localisation~\cite{anderson1958absence,lee1985disordered} in non-interacting and many-body localisation~\cite{nandkishore2015many,alet2018many,abanin2019colloquium} in interacting systems.

However, disorder is not a prerequsite for ergodicity breaking. In translation-invariant systems, kinetic constraints have long emerged as one of the most prominent pathways to classical glassy behaviour at low temperatures~\cite{fredrickson1984kinetic,fredrickson1985facilitated,jackle1991hierarchically,sollich1999glassy,sollich2003glassy,ritort2003glassy,garrahan2011kinetically}. Also in quantum systems, at infinite temperatures, kinetic constraints have been shown to result in slow relaxation~\cite{horssen2015dynamics,lan2018quantum,pancotti2020quantum} and stabilise a many-body localised phase as well~\cite{roy2020strong}.
This motivates our investigation of the fate of classical many-body chaos, characterised by the classical OTOC, in the presence of local kinetic constraints and at infinite temperatures.

Remarkably we find that, as a function of a parameter $\phi$ quantifying the strength of the constraints, the dynamics undergoes a sharp `dynamical phase transition' between an ergodic phase, where the OTOC spreads out ballistically, and a non-ergodic phase where spreading is completely arrested--see Fig.~\ref{fig:otocs} for a summary. As the constraints in our model keep the entire configuration space dynamically connected, localisation is \emph{not} due to fragmentation of configuration space~\cite{pai2019localisation,khemani2020localisation,sala2020ergodicity,feldmeir2021critically}.  

\begin{figure}[t]
  \includegraphics[width=\linewidth]{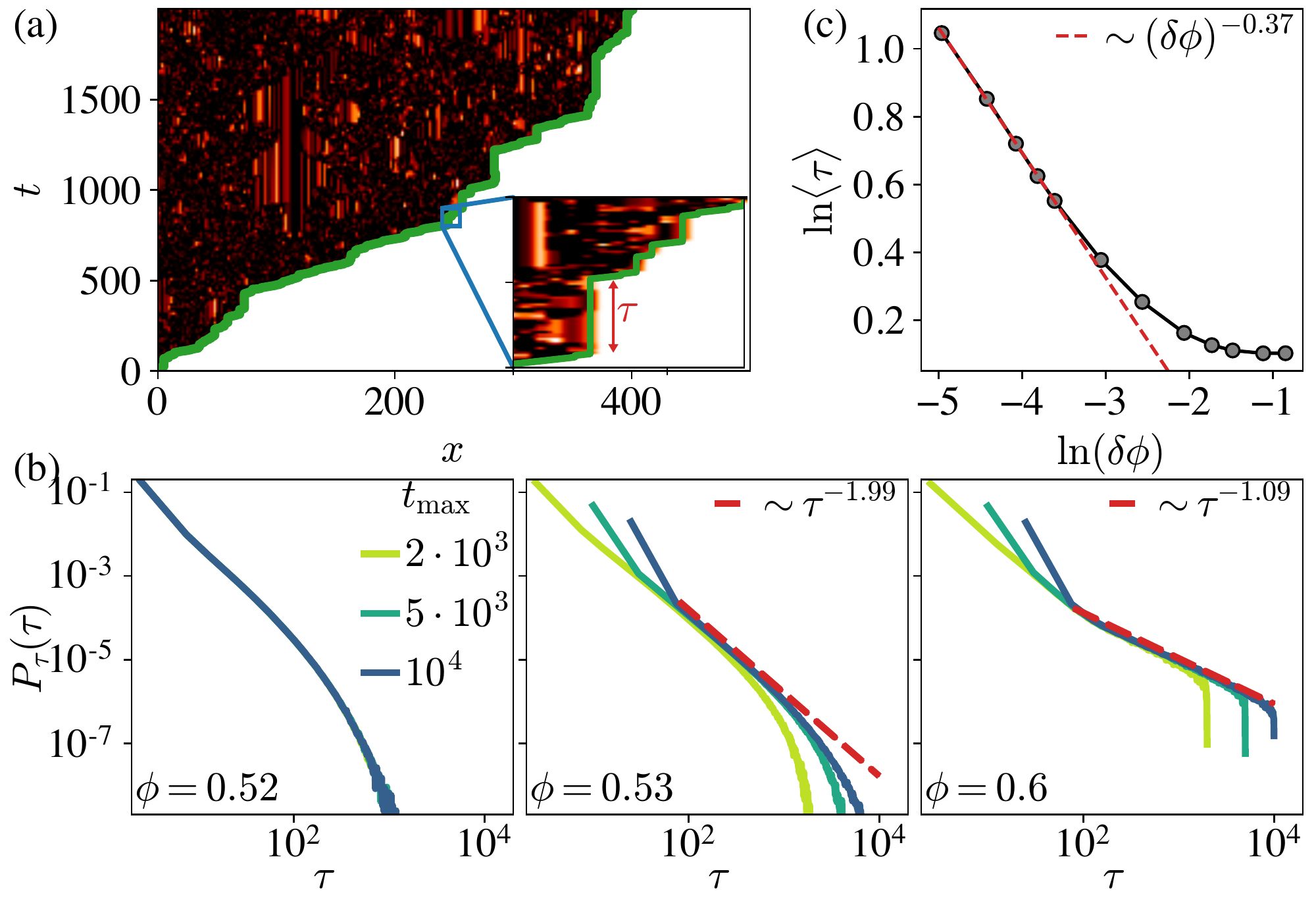}
  \caption{(a) $D(x,t)$ for a randomly chosen initial condition along with the trajectory of the front, $x_F^R(t)$, extracted using Eq.~\eqref{eq:front}, shown in green. The inset shows a small spacetime portion zoomed in as well as the definition of $\tau$. Data for $\phi=0.53$ and $\eta=0.01$. (b) Distributions of the waiting time $P_\tau(\tau)$ for three different $\phi$, each for three different maximum simulation time, $t_\mathrm{max}$. While the distributions are converged with $t_\mathrm{max}$ and not tailed in the delocalised phase, they have heavy power-law tails in the localised phase which persist for longer $\tau$ for larger $t_\mathrm{max}$ and have exponents such that the mean $\braket{\tau}$ is divergent. (c) The numerical data suggests that $\braket{\tau}$ diverges as a power-law with $\delta\phi=\phi_c-\phi$ and $\phi_c\approx 0.525(5)$. All statistics accumulated over $5\times 10^5$ initial conditions.}
    \label{fig:ptau-tau}
\end{figure}

\begin{figure}[!b]
\includegraphics[width=\linewidth]{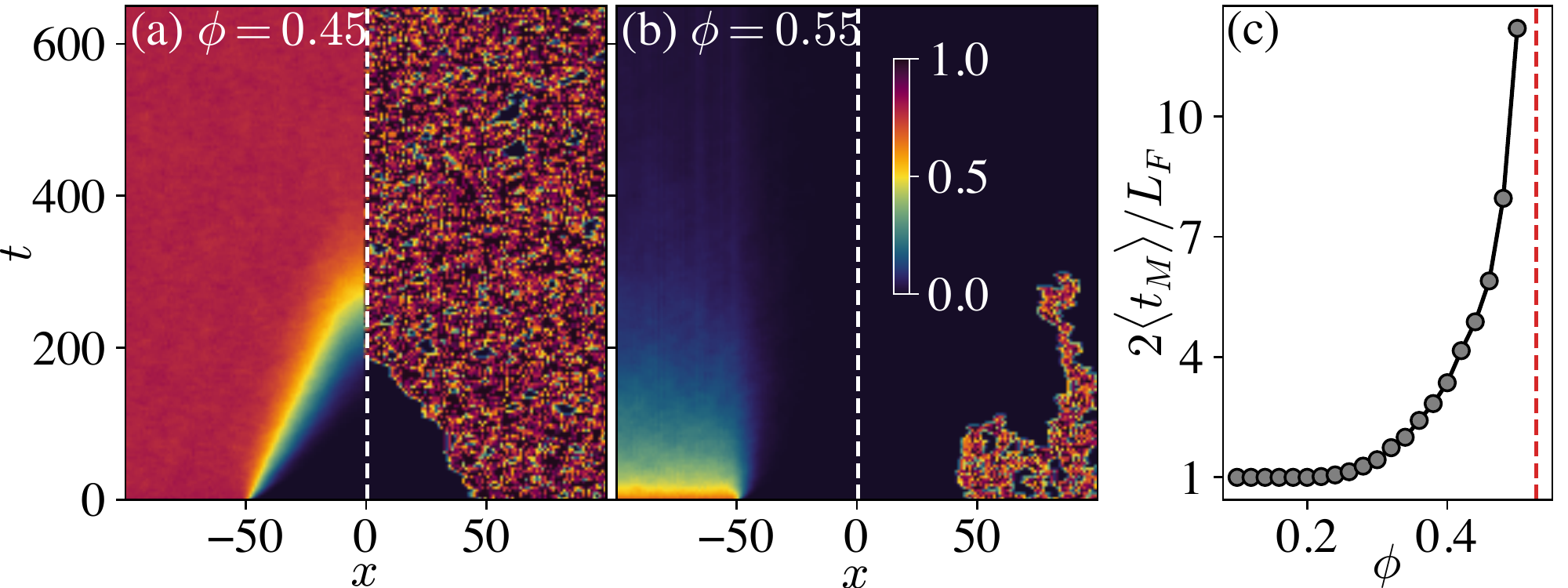}
\caption{Melting of an initially frozen island of length $L_F=100$ in the delocalised phase (a) or lack thereof in the localised phase (b). The plots show a colourmap of $A_x(t)$ with black denoting frozen spins $A_x(t)\approx0$ and yellow, active with $A_x(t)\approx1$. In each panel, the left half shows averaged data, $\braket{A(x,t)}$, whereas the right half shows it for a single initial condition. The data in (c) suggests that the mean melting time $t_M$, diverges as the critical $\phi_c$ is approached from the delocalised side. The red dashed line denotes the estimated $\phi_c$ from the decorrelator data.}
\label{fig:melting}
\end{figure}

Instead, localisation occurs because the distribution of waiting times (the time the front of the classical OTOC waits at a site before moving onto the next) above some critical constraint strength acquires a heavy enough power-law tail that the mean waiting time diverges (Fig.~\ref{fig:ptau-tau}). We link this distribution to the broadening of the front caused by the constraints, and provide insight into the mechanism of localisation by showing that melting of initially frozen regions happens locally, starting at the edges of the islands, while formation of frozen islands can occur anywhere in the system. Whether the entire system becomes localised is then a question of the lifetimes of these islands, which diverges at the same $\phi$ as the mean waiting time (Fig.~\ref{fig:melting}).

Further evidence for localisation is provided by the fact that, in the ergodic phase, islands of initially frozen spins (due to the constraints) melt from the edges and eventually the entire system becomes active. In the localised phase, not only is melting of these initial frozen islands arrested, but the system also dynamically develops several frozen islands with divergent lifetimes, which eventually proliferate and freeze the entire system.

For simplicity, we strip our model of all conservation laws (including energy) so that in the unconstrained limit ($\phi=0$), the OTOC has a sharp front with no broadening. In the presence of constraints but in the ergodic phase the distribution of waiting times broadens the front diffusively.

For concreteness, we consider a periodically-driven classical Heisenberg chain of length $L$ with periodic boundaries described by the Hamiltonian,

\begin{equation}
  \mathcal{H}(t)=
  \begin{cases}
  \sum\limits_{x=-L/2}^{L/2} (J \s{z}{x}\s{z}{x+1}+h \s{z}{x}),~t \in[nT, (n+\frac{1}{2})T)\\
  \sum\limits_{x=-L/2}^{L/2}  g \s{x}{x}\,,~~t\in [(n+\frac{1}{2})T,(n+1)T)
  \end{cases}\,,
\label{eq:hamiltonian}
\end{equation}
where the spin at site $x$ is a three-dimensional unit vector, $\mathbf{S}_x=(S_x^x,S_x^y,S_x^z)$.
Since the model defined in Eq.~\eqref{eq:hamiltonian} is time-periodic we consider the dynamics only at stroboscopic times $t=n T$ with integer $n$. In the presence of kinetic constraints, the stroboscopic evolution of the spins are given by 
\begin{equation}
\mathbf{S}_x[(n+1)T] = \begin{cases} \mathsf{R_x}[gT/2]\mathsf{R_z}[\theta_x(nT)] \mathbf{S}_x(nT);~\Theta_x(nT)=1\\
\mathbf{S}_x(nT);~\Theta_x(nT)=0
\end{cases}\,,
\end{equation}
where $\mathsf{R_z}[\theta_x(nT)]$ denotes a 3D rotation matrix about the $z$-axis by an angle $\theta_{x}(nT)\equiv [S^z_{x-1}(nT)+S^z_{x+1}(nT)+h]T/2$ and similarly for $\mathsf{R_x}$.
The constraints are encoded in the Heaviside step function,
\begin{equation}
\Theta_x(nT)=\Theta[\cos(\pi\phi)-\min(S_{x-1}^z(nT),S_{x+1}^z(nT))]\,.
\label{eq:constraints}
\end{equation}
The form of the constraint implies that a spin at site $x$ is frozen if \emph{both} its neighbours lie inside the spherical sector defined by the polar angle $\pi\phi$, see Fig.~\ref{fig:otocs}(a) for a visual schematic. In this sense, it can be considered as the Heisenberg generalisation of the Fredrickson-Andersen (FA) constraint defined originally for Ising spins~\cite{fredrickson1984kinetic,fredrickson1985facilitated}. Physically, in the context of glassy dynamics, interpreting spins inside (outside) the sector as proxies for (for example) high (low)-density regions in a system where density is the dynamical variable, the constraints forbid dynamics in a region if it is surrounded by dense immobile regions~\cite{ritort2003glassy}.
In what follows we set $T=2\pi$, $J=1$, $h=0.1$, and $g=0.4$ without loss of generality. 

The classical OTOC is defined by considering two initial conditions $\{\mathbf{S}_{x,A}(t=0)\}$ and $\{\mathbf{S}_{x,B}(t=0)\}$  identical everywhere except at $x=0$, where they are infinitesimally different:
\begin{equation}
  \delta\mathbf{S}_x(0) \equiv \mathbf{S}_{x,A}(0)-\mathbf{S}_{x,B}(0) = \varepsilon\delta_{x,0}[\hat{\mathbf{z}}\times \mathbf{S}_{x,A}(0)]\,.
\end{equation}
The classical OTOC, henceforth referred to as the \emph{decorrelator}, is then given by~\cite{das2018light}
\begin{equation}
D(x,t) = 1-\mathbf{S}_{x,A}(t)\cdot\mathbf{S}_{x,B}(t)\,.
\label{eq:D}
\end{equation}
As we are interested in the infinite temperature dynamics of the decorrelator, we average it over several randomly and uniformly chosen initial conditions; we denote the average as $\braket{D(x,t)}$~\footnote{As shown in Ref.~\cite{das2018light}, on canonical quantisation of the Poisson brackets $\braket{D(x,t)}\to -(\varepsilon^2/\hbar^2)\mathrm{Tr}[[\hat{\mathbf{S}}_x(t),\hat{\mathbf{z}}\cdot \hat{\mathbf{S}}_0(0)]^2]$ which is nothing but the quantum OTOC.}, and use $\varepsilon=10^{-3}$ throughout. 

The results for $\braket{D(x,t)}$ are shown in Fig.~\ref{fig:otocs}(b) for several values of $\phi$. For $\phi=0$, the dynamics is completely unconstrained. Since the system has no conservation laws, we observe a sharp ballistic lightcone for $\braket{D(x,t)}$ the front of which does not broaden. On increasing $\phi$, the constraints come into play and within the `delocalised' phase while the lightcone still has a well-defined butterfly velocity, there is broadening of the front. We shall shortly explain this using the distribution of waiting times. In the vicinity of $\phi_c\approx 0.53$, the lightcone slows down while the front broadens significantly. We attribute this slow dynamics with large fluctuations to the fact that we are at or near a `dynamical phase transition' separating the delocalised and localised phases. The localised phase is evident from the data for $\phi=0.6$ where the lightcone is completely arrested; the front neither propagates nor broadens.

A key ingredient that determines the spatiotemporal profile of the decorrelator is the \emph{waiting time} denoted by $\tau$. This is defined as the time the \emph{front} of the decorrelator waits at $x$ before moving to site $x+\mathrm{sgn}(x)$. The front of the decorrelator is defined as follows. For a fixed time $t$, the right/left front, $x_F^{(R/L)}(t)$ is at
\begin{equation}
x_F^{(R/L)}(t) = \max/\min\{x | D(x,t)\ge \eta\}\,,
\label{eq:front}
\end{equation}
where $\eta \ll 1$ is an empirically chosen cutoff. A representative trajectory of $x_F^R$ over time is shown in Fig.~\ref{fig:ptau-tau}(a), where a series of vertical steps (the length of each being a waiting time, $\tau$) is visible. These steps are distributed over spacetime and initial conditions with a distribution $P_\tau$, shown in Fig.~\ref{fig:ptau-tau}(b) for various $\phi$ and simulation times.
In the delocalised phase $(\phi=0.52)$, $P_\tau$ converges with increasing maximum simulation time $t_\mathrm{max}$, decaying faster than any power-law and having a finite mean $\braket{\tau}=\int^\infty d\tau~\tau P_\tau(\tau)$. As the system transitions into the localised phase ($\phi=0.53$ and $\phi=0.6$), $P_\tau(\tau)$ develops power-law tails, $P_\tau(\tau)\sim\tau^{-\alpha}$. These tails are cut off by $t_\mathrm{max}$ but, crucially, extend up to longer $\tau$ for larger $t_\mathrm{max}$ This suggests that when $t_\mathrm{max}\to\infty$ the tail remains a power law all the way to $\tau\to\infty$. For $\phi\ge \phi_c \approx 0.525(5)$, the exponent $\alpha \le 2$, {\textit{i.e.}} the tail of $P_\tau$ becomes heavy enough that $\braket{\tau}\to\infty$. From the data in Fig.~\ref{fig:ptau-tau}(b)., we conjecture that at $\phi=\phi_c$, $\alpha=2$
 The time taken by the front to reach a site at a given $x$ is $\sim x\braket{\tau}$, which diverges so that the front gets stuck  and the lightcone does not spread at all \footnote{Strictly speaking, this divergence rules out the ballistic spreading of the front. Subsequent results on melting of an initially frozen island (Fig.~\ref{fig:melting}) provides evidence for the complete arrest of the lightcone.}.
In Fig.~\ref{fig:ptau-tau}(c) we show how $\braket{\tau}$ diverges as $\phi\to\phi_c$ from the delocalised side. The data on logarithmic axes shows that it diverges as a power-law with an exponent $\nu\approx 0.4$~\footnote{A systematic study of the critical behaviour incorporating finite-size and finite-time effects is beyond the scope and interest of this work.}.

So far we have established that the constraints \eqref{eq:constraints} induce a `localised' phase wherein classical many-body chaos as quantified by the decorrelator, \eqref{eq:D}, is completely arrested. We next show that in fact the entire system actually freezes in a random spin configuration which depends on the initial condition, and provide a physical picture for localisation.

The mechanism can be understood in terms of  \emph{frozen islands}: contiguous segments of spins such that $\Theta_x=0$ (see Eq.~\eqref{eq:constraints}) for all of them. All spins in such a segment and the spins immediately on either side have $S^z>\cos(\pi\phi)$, hence are frozen. During the evolution such islands can melt (the spins becoming active) progressively inwards from the edges of the island.  In the delocalised phase, these islands melt at a finite velocity and the entire system becomes active at late times. In the localised phase new frozen islands appear, proliferating through the entire system such that it freezes into a random configuration. Of course, for any value of $\phi$, new frozen islands appear dynamically. However, the crucial difference is that, for $\phi<\phi_c$, these islands quickly melt with timescales that are proportional to the length of the island. On the contrary, for $\phi>\phi_c$, the rate at which new islands appear overwhelms the rate at which they melt such that the entire system becomes one frozen island. 

We next present numerical evidence for this picture. We start with an ensemble of initial conditions with a frozen island of size $L_F$ in the middle and the rest of the spins random. We then define a function $A_x(t) = \Theta[\cos(\pi\phi)-\min(S_{x-1}^z(t),S_{x+1}^z(t))]$, taking a value 1(0) if the spin is active(frozen), and track this dynamically. The results are shown in Fig.~\ref{fig:melting}, and are in agreement with the discussion in the previous paragraph. Furthermore, from the data for a single initial condition, it can be seen that new frozen islands form dynamically in the initially active regions, quickly melt in the delocalised phase, and persist and eventually take over the entire system in the localised phase.

To quantify this further, we define a melting time, $t_M$, for each initial condition as the earliest time the spin at $x=0$ (furthest from the edges of the island) becomes active, $t_M = \min\{t| A_0(t)=1\}$. In the delocalised phase, since the melting happens at the constant velocity on average, we expect that the average melting time $\braket{t_M}\propto L_F$. However, in the localised phase, since the initial frozen island never melts, the spin at site $x=0$ is never active for any finite $t$. This is consistent with the apparent divergence of $\braket{t_M}$ as $\phi\to\phi_c$ from the delocalised side in Fig.~\ref{fig:melting}(c).

\begin{figure}[t]
  \includegraphics[width=1\columnwidth]{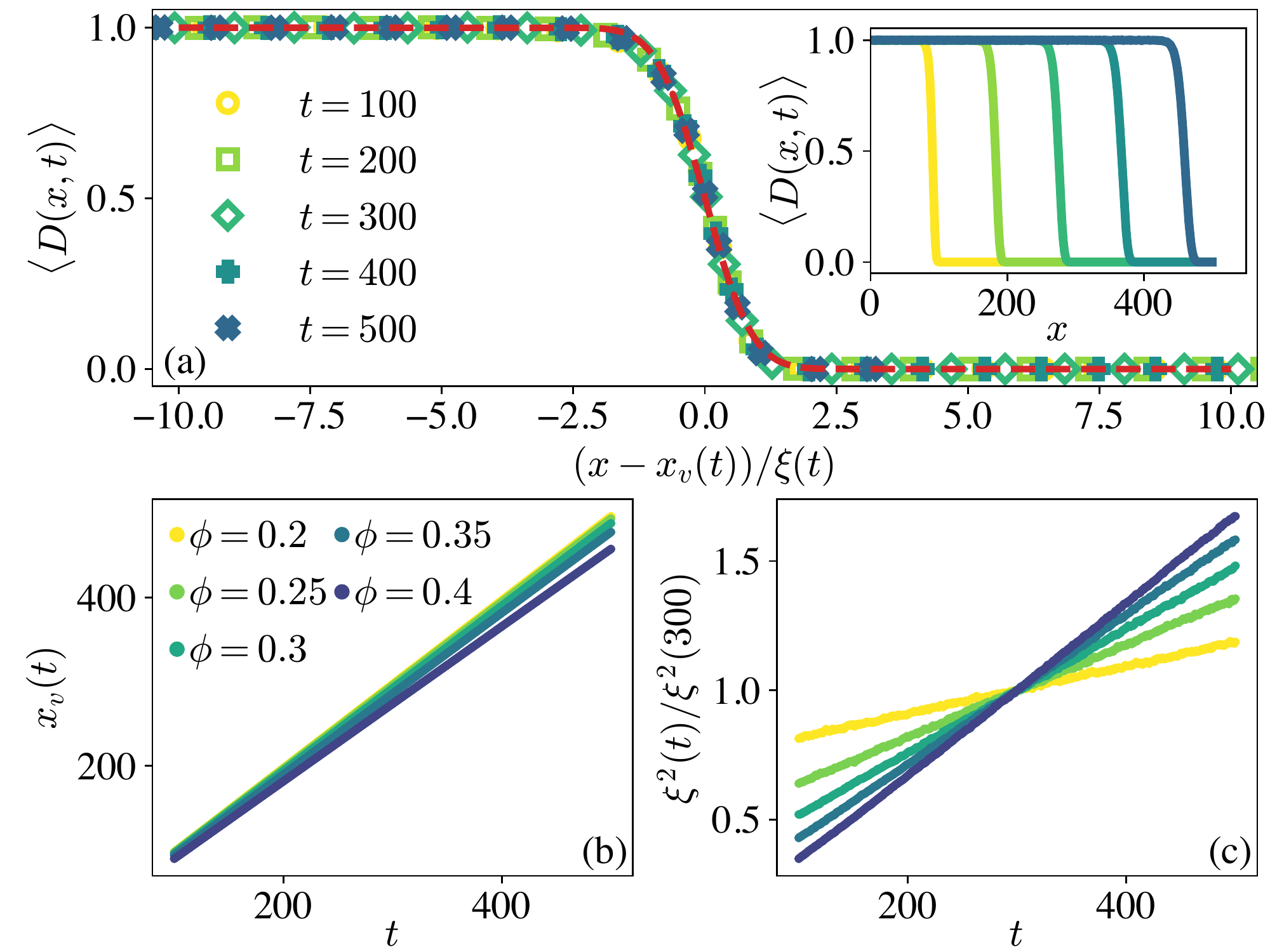}
  \caption{
    (a) Collapse of $\braket{D(x,t)}$ onto a function $\mathcal{F}[(x-x_v(t))/\xi(t)]$ for $x>0$ with $\mathcal{F}(y)=\erfc(y)$ denoted by the red dashed line. The inset shows the raw data. The plots use $\phi=0.4$. (b) $x_v(t)$ as a function of $t$ for several values of $\phi$ in the delocalised phase. The linear behaviour indicates the presence of a well-defined butterfly velocity $v_B = x_v(t)/t$. (c) $\xi^2(t)$ for the same values of $\phi$, rescaled (arbitrarily) with  $\xi^2(300)$ for visual clarity. The linear behaviour indicates the diffusive broadening of the decorrelator front.}
    \label{fig:broad}
\end{figure}

Finally, we discuss the spatiotemporal profile of the decorrelator in the delocalised phase focussing in particular on the broadening of the front. For $\phi=0$ there is no broadening so that any broadening at $\phi\neq 0$ must be due to the constraints. 
As the top panel in Fig.~\ref{fig:broad} demonstrates for a representative value of $\phi$, the mean decorrelator for different times can be collapsed onto a scaling function,
\begin{equation}
\braket{D(x,t)} = \mathcal{F}\left[\frac{x \mp x_v(t)}{\xi(t)}\right]\,; x\gtrless 0\,.
\end{equation}
Fitting for each $t$ (lower two panels) shows that the parameters $x_v(t)= v_B(\phi) t$ and $\xi(t)= \gamma(\phi) t^{1/2}$ so that the front moves ballistically while broadening diffusively. The broadening is a direct result of the finite width of the distribution $P_\tau$. Defining $\ell$ as the distance the front moves in a single period, the finite width of $P_\tau$ naturally implies that the distribution $P_\ell$ has a non-zero width. Moreover, since the model is local, $\ell$ is strictly bounded from above which means all the moments of $P_\ell$ are finite.
Let us denote the first two moments by $\pm\mu_\ell$ (for the front at $x\gtrless 0$ and with $\mu_\ell>0$) and $\sigma_\ell^2$, respectively. Acccording to the central limit theorem the distance moved after $t$ periods, $X_t=\sum_{n=1}^t \ell_n$ with $\ell_n$ distributed according to $P_\ell$, is normally distributed with mean $t\mu_\ell$ and standard deviation $\sqrt{t}\sigma_\ell$. Modelling the decorrelator for a single initial condition by a Heaviside step function, $D(x,t)=\Theta(X_t-|x|)$, and averaging over the normal distribution of the $X_t$, we find 
\begin{equation}
\braket{D(x,t)} = \frac{1}{2}\erfc\left(\frac{x\mp t\mu_\ell}{\sqrt{2t\sigma^2_\ell}}\right)\,; x\gtrless 0\,,
\label{eq:erf}
\end{equation}
so that $v_B(\phi)= \mu_\ell(\phi)$ and $\gamma(\phi)=\sqrt{2}\sigma_\ell(\phi)$.
An implication of Eq.~\eqref{eq:erf} is that on a spacetime ray with velocity $v$ ($x=vt$) outside the lightcone we can write $\braket{D(x,t)}\sim \exp[\lambda_v t]$ where $\lambda_v(\phi)\approx- [v-v_B(\phi)]^2/\gamma(\phi)$ is the velocity-dependent Lyapunov exponent~\cite{khemani2018velocity}.

To summarise, we have demonstrated that constrained dynamics can completely arrest classical many-body chaos as measured via the classical counterpart of the OTOC. Our results also provide compelling evidence for a `dynamical phase transition', driven by the strength of the constraint, separating a delocalised phase where the classical OTOC spreads ballistically from a localised phase where it does not spread at all. The physical mechanism behind this localisation was shown to be that, in the course of the dynamics, frozen islands of spins form and proliferate through the system, freezing it entirely. These islands also form dynamically in the delocalised phase but they quickly melt away. A consequence of the constraint-induced frozen spins is that as correlations spread in spacetime, they encounter the frozen spins and need to wait until they are dynamically active again. The arrest of spreading is reflected in the distribution of these waiting times acquiring heavy power-law tails in the localised phase.

It is worth emphasising that our results are general, holding for a variety of systems with and without conservation laws and both in one and two dimensions~\cite{supp}. The form of the constraint we employ, \eqref{eq:constraints}, allows for a spin to be active even if just one of its neighbours is not in the cone (\textit{i.e.}, OR condition). In this sense, the constraint is weaker than the generalisations of XOR or AND constraints where exactly one of the two or both neighbours, respectively, need to be outside the cone for a spin to be active. This therefore suggests that the localised phase will be present for the latter two cases.

While we have established the presence of a transition between an ergodic and a localised phase, its precise nature is a question for future work. In a separate work~\cite{deger2022phase} we discuss how it can be mapped onto a directed percolation problem. 

Looking further afield, one may ask what lessons can be learnt from this classical problem for constrained quantum dynamics. Amidst the fragility of many-body localised phases in disordered quantum systems, particularly in higher dimensions, the emergence of constraints as an effective ingredient for localising quantum information is significant. This aspect also has important practical implications. Modern day noisy intermediate scale quantum devices simulate dynamics to store, manipulate and retrieve quantum information~\cite{preskill2018quantum}. An essential challenge there is scrambling of information accompanied by runaway entropy growth, or heating, of the system. Constraints as a way of mitigating this issue without relying on disorder to break ergodicity is an important and timely development. Our results point to a clear direction for further studies on how to arrest quantum chaos leading to information localisation.

\begin{acknowledgments}
We thank Y. Bar Lev, S. Bhattacharjee and A. Dhar for helpful discussions, and A. Smith for a careful reading of the manuscript. This work was in part supported by EPSRC Grant No. EP/V012177/1 (AD and AL), ICTS-Simons Early Career Faculty Fellowship (SR) and EPSRC Grant No. EP/S020527/1 (SR).
\end{acknowledgments}

\bibliography{references}

\begin{thebibliography}{53}%
\makeatletter
\providecommand \@ifxundefined [1]{%
 \@ifx{#1\undefined}
}%
\providecommand \@ifnum [1]{%
 \ifnum #1\expandafter \@firstoftwo
 \else \expandafter \@secondoftwo
 \fi
}%
\providecommand \@ifx [1]{%
 \ifx #1\expandafter \@firstoftwo
 \else \expandafter \@secondoftwo
 \fi
}%
\providecommand \natexlab [1]{#1}%
\providecommand \enquote  [1]{``#1''}%
\providecommand \bibnamefont  [1]{#1}%
\providecommand \bibfnamefont [1]{#1}%
\providecommand \citenamefont [1]{#1}%
\providecommand \href@noop [0]{\@secondoftwo}%
\providecommand \href [0]{\begingroup \@sanitize@url \@href}%
\providecommand \@href[1]{\@@startlink{#1}\@@href}%
\providecommand \@@href[1]{\endgroup#1\@@endlink}%
\providecommand \@sanitize@url [0]{\catcode `\\12\catcode `\$12\catcode
  `\&12\catcode `\#12\catcode `\^12\catcode `\_12\catcode `\%12\relax}%
\providecommand \@@startlink[1]{}%
\providecommand \@@endlink[0]{}%
\providecommand \url  [0]{\begingroup\@sanitize@url \@url }%
\providecommand \@url [1]{\endgroup\@href {#1}{\urlprefix }}%
\providecommand \urlprefix  [0]{URL }%
\providecommand \Eprint [0]{\href }%
\providecommand \doibase [0]{https://doi.org/}%
\providecommand \selectlanguage [0]{\@gobble}%
\providecommand \bibinfo  [0]{\@secondoftwo}%
\providecommand \bibfield  [0]{\@secondoftwo}%
\providecommand \translation [1]{[#1]}%
\providecommand \BibitemOpen [0]{}%
\providecommand \bibitemStop [0]{}%
\providecommand \bibitemNoStop [0]{.\EOS\space}%
\providecommand \EOS [0]{\spacefactor3000\relax}%
\providecommand \BibitemShut  [1]{\csname bibitem#1\endcsname}%
\let\auto@bib@innerbib\@empty
\bibitem [{\citenamefont {Landau}\ and\ \citenamefont
  {Lifshitz}(1980)}]{landaulifshitz}%
  \BibitemOpen
  \bibfield  {author} {\bibinfo {author} {\bibfnamefont {L.~D.}\ \bibnamefont
  {Landau}}\ and\ \bibinfo {author} {\bibfnamefont {E.~M.}\ \bibnamefont
  {Lifshitz}},\ }\href
  {https://doi.org/https://doi.org/10.1016/C2009-0-24487-4} {\emph {\bibinfo
  {title} {Statistical Physics: Vol. 5}}},\ \bibinfo {edition} {3rd}\ ed.\
  (\bibinfo  {publisher} {Butterworth-Heinemann},\ \bibinfo {address}
  {Oxford},\ \bibinfo {year} {1980})\BibitemShut {NoStop}%
\bibitem [{\citenamefont {Lorenz}(1963)}]{lorenz1963deterministic}%
  \BibitemOpen
  \bibfield  {author} {\bibinfo {author} {\bibfnamefont {E.~N.}\ \bibnamefont
  {Lorenz}},\ }\bibfield  {title} {\bibinfo {title} {Deterministic nonperiodic
  flow},\ }\href {https://doi.org/10.1175/1520-0469(1963)020<0130:DNF>2.0.CO;2}
  {\bibfield  {journal} {\bibinfo  {journal} {Journal of Atmospheric Sciences}\
  }\textbf {\bibinfo {volume} {20}},\ \bibinfo {pages} {130 } (\bibinfo {year}
  {1963})}\BibitemShut {NoStop}%
\bibitem [{\citenamefont {Lorenz}(1995)}]{lorenz1995}%
  \BibitemOpen
  \bibfield  {author} {\bibinfo {author} {\bibfnamefont {E.~N.}\ \bibnamefont
  {Lorenz}},\ }\href
  {https://uwapress.uw.edu/book/9780295975146/the-essence-of-chaos} {\emph
  {\bibinfo {title} {The Essence of Chaos}}}\ (\bibinfo  {publisher}
  {University of Washington Press},\ \bibinfo {address} {Seattle},\ \bibinfo
  {year} {1995})\BibitemShut {NoStop}%
\bibitem [{\citenamefont {Lorenz}(2001)}]{lorenz2001butterfly}%
  \BibitemOpen
  \bibfield  {author} {\bibinfo {author} {\bibfnamefont {E.}~\bibnamefont
  {Lorenz}},\ }\bibinfo {title} {The {B}utterfly {E}ffect},\ in\ \href
  {https://doi.org/10.1142/9789812386472_0007} {\emph {\bibinfo {booktitle}
  {The Chaos Avant-Garde}}}\ (\bibinfo  {publisher} {World Scientific},\
  \bibinfo {address} {Singapore},\ \bibinfo {year} {2001})\ pp.\ \bibinfo
  {pages} {91--94}\BibitemShut {NoStop}%
\bibitem [{\citenamefont {{Larkin}}\ and\ \citenamefont
  {{Ovchinnikov}}(1969)}]{larkin1969quasisclassical}%
  \BibitemOpen
  \bibfield  {author} {\bibinfo {author} {\bibfnamefont {A.~I.}\ \bibnamefont
  {{Larkin}}}\ and\ \bibinfo {author} {\bibfnamefont {Y.~N.}\ \bibnamefont
  {{Ovchinnikov}}},\ }\bibfield  {title} {\bibinfo {title} {{Quasiclassical
  Method in the Theory of Superconductivity}},\ }\href@noop {} {\bibfield
  {journal} {\bibinfo  {journal} {Soviet Journal of Experimental and
  Theoretical Physics}\ }\textbf {\bibinfo {volume} {28}},\ \bibinfo {pages}
  {1200} (\bibinfo {year} {1969})}\BibitemShut {NoStop}%
\bibitem [{\citenamefont {Shenker}\ and\ \citenamefont
  {Stanford}(2014)}]{shenker2014black}%
  \BibitemOpen
  \bibfield  {author} {\bibinfo {author} {\bibfnamefont {S.~H.}\ \bibnamefont
  {Shenker}}\ and\ \bibinfo {author} {\bibfnamefont {D.}~\bibnamefont
  {Stanford}},\ }\bibfield  {title} {\bibinfo {title} {Black holes and the
  butterfly effect},\ }\href {http://dx.doi.org/10.1007/jhep03(2014)067}
  {\bibfield  {journal} {\bibinfo  {journal} {Journal of High Energy Physics}\
  }\textbf {\bibinfo {volume} {2014}},\ \bibinfo {pages} {067} (\bibinfo {year}
  {2014})}\BibitemShut {NoStop}%
\bibitem [{\citenamefont {Roberts}\ and\ \citenamefont
  {Stanford}(2015)}]{roberts2015diagnosing}%
  \BibitemOpen
  \bibfield  {author} {\bibinfo {author} {\bibfnamefont {D.~A.}\ \bibnamefont
  {Roberts}}\ and\ \bibinfo {author} {\bibfnamefont {D.}~\bibnamefont
  {Stanford}},\ }\bibfield  {title} {\bibinfo {title} {Diagnosing chaos using
  four-point functions in two-dimensional conformal field theory},\ }\href
  {https://doi.org/10.1103/PhysRevLett.115.131603} {\bibfield  {journal}
  {\bibinfo  {journal} {Phys. Rev. Lett.}\ }\textbf {\bibinfo {volume} {115}},\
  \bibinfo {pages} {131603} (\bibinfo {year} {2015})}\BibitemShut {NoStop}%
\bibitem [{\citenamefont {Maldacena}\ \emph {et~al.}(2016)\citenamefont
  {Maldacena}, \citenamefont {Shenker},\ and\ \citenamefont
  {Stanford}}]{maldacena2016bound}%
  \BibitemOpen
  \bibfield  {author} {\bibinfo {author} {\bibfnamefont {J.}~\bibnamefont
  {Maldacena}}, \bibinfo {author} {\bibfnamefont {S.~H.}\ \bibnamefont
  {Shenker}},\ and\ \bibinfo {author} {\bibfnamefont {D.}~\bibnamefont
  {Stanford}},\ }\bibfield  {title} {\bibinfo {title} {A bound on chaos},\
  }\href {http://dx.doi.org/10.1007/JHEP08(2016)106} {\bibfield  {journal}
  {\bibinfo  {journal} {Journal of High Energy Physics}\ }\textbf {\bibinfo
  {volume} {2016}},\ \bibinfo {pages} {106} (\bibinfo {year}
  {2016})}\BibitemShut {NoStop}%
\bibitem [{\citenamefont {Swingle}\ \emph {et~al.}(2016)\citenamefont
  {Swingle}, \citenamefont {Bentsen}, \citenamefont {Schleier-Smith},\ and\
  \citenamefont {Hayden}}]{swingle2016measuring}%
  \BibitemOpen
  \bibfield  {author} {\bibinfo {author} {\bibfnamefont {B.}~\bibnamefont
  {Swingle}}, \bibinfo {author} {\bibfnamefont {G.}~\bibnamefont {Bentsen}},
  \bibinfo {author} {\bibfnamefont {M.}~\bibnamefont {Schleier-Smith}},\ and\
  \bibinfo {author} {\bibfnamefont {P.}~\bibnamefont {Hayden}},\ }\bibfield
  {title} {\bibinfo {title} {Measuring the scrambling of quantum information},\
  }\href {https://doi.org/10.1103/PhysRevA.94.040302} {\bibfield  {journal}
  {\bibinfo  {journal} {Phys. Rev. A}\ }\textbf {\bibinfo {volume} {94}},\
  \bibinfo {pages} {040302} (\bibinfo {year} {2016})}\BibitemShut {NoStop}%
\bibitem [{\citenamefont {Bohrdt}\ \emph {et~al.}(2017)\citenamefont {Bohrdt},
  \citenamefont {Mendl}, \citenamefont {Endres},\ and\ \citenamefont
  {Knap}}]{bohrdt2017scrambling}%
  \BibitemOpen
  \bibfield  {author} {\bibinfo {author} {\bibfnamefont {A.}~\bibnamefont
  {Bohrdt}}, \bibinfo {author} {\bibfnamefont {C.~B.}\ \bibnamefont {Mendl}},
  \bibinfo {author} {\bibfnamefont {M.}~\bibnamefont {Endres}},\ and\ \bibinfo
  {author} {\bibfnamefont {M.}~\bibnamefont {Knap}},\ }\bibfield  {title}
  {\bibinfo {title} {Scrambling and thermalization in a diffusive quantum
  many-body system},\ }\href {https://doi.org/10.1088/1367-2630/aa719b}
  {\bibfield  {journal} {\bibinfo  {journal} {New Journal of Physics}\ }\textbf
  {\bibinfo {volume} {19}},\ \bibinfo {pages} {063001} (\bibinfo {year}
  {2017})}\BibitemShut {NoStop}%
\bibitem [{\citenamefont {Luitz}\ and\ \citenamefont
  {Bar~Lev}(2017)}]{luitz2017information}%
  \BibitemOpen
  \bibfield  {author} {\bibinfo {author} {\bibfnamefont {D.~J.}\ \bibnamefont
  {Luitz}}\ and\ \bibinfo {author} {\bibfnamefont {Y.}~\bibnamefont
  {Bar~Lev}},\ }\bibfield  {title} {\bibinfo {title} {Information propagation
  in isolated quantum systems},\ }\href
  {https://doi.org/10.1103/PhysRevB.96.020406} {\bibfield  {journal} {\bibinfo
  {journal} {Phys. Rev. B}\ }\textbf {\bibinfo {volume} {96}},\ \bibinfo
  {pages} {020406} (\bibinfo {year} {2017})}\BibitemShut {NoStop}%
\bibitem [{\citenamefont {Nahum}\ \emph {et~al.}(2018)\citenamefont {Nahum},
  \citenamefont {Vijay},\ and\ \citenamefont {Haah}}]{nahum2018operator}%
  \BibitemOpen
  \bibfield  {author} {\bibinfo {author} {\bibfnamefont {A.}~\bibnamefont
  {Nahum}}, \bibinfo {author} {\bibfnamefont {S.}~\bibnamefont {Vijay}},\ and\
  \bibinfo {author} {\bibfnamefont {J.}~\bibnamefont {Haah}},\ }\bibfield
  {title} {\bibinfo {title} {Operator spreading in random unitary circuits},\
  }\href {https://doi.org/10.1103/PhysRevX.8.021014} {\bibfield  {journal}
  {\bibinfo  {journal} {Phys. Rev. X}\ }\textbf {\bibinfo {volume} {8}},\
  \bibinfo {pages} {021014} (\bibinfo {year} {2018})}\BibitemShut {NoStop}%
\bibitem [{\citenamefont {von Keyserlingk}\ \emph {et~al.}(2018)\citenamefont
  {von Keyserlingk}, \citenamefont {Rakovszky}, \citenamefont {Pollmann},\ and\
  \citenamefont {Sondhi}}]{keyserlingk2018operator}%
  \BibitemOpen
  \bibfield  {author} {\bibinfo {author} {\bibfnamefont {C.~W.}\ \bibnamefont
  {von Keyserlingk}}, \bibinfo {author} {\bibfnamefont {T.}~\bibnamefont
  {Rakovszky}}, \bibinfo {author} {\bibfnamefont {F.}~\bibnamefont
  {Pollmann}},\ and\ \bibinfo {author} {\bibfnamefont {S.~L.}\ \bibnamefont
  {Sondhi}},\ }\bibfield  {title} {\bibinfo {title} {Operator hydrodynamics,
  otocs, and entanglement growth in systems without conservation laws},\ }\href
  {https://doi.org/10.1103/PhysRevX.8.021013} {\bibfield  {journal} {\bibinfo
  {journal} {Phys. Rev. X}\ }\textbf {\bibinfo {volume} {8}},\ \bibinfo {pages}
  {021013} (\bibinfo {year} {2018})}\BibitemShut {NoStop}%
\bibitem [{\citenamefont {Rakovszky}\ \emph {et~al.}(2018)\citenamefont
  {Rakovszky}, \citenamefont {Pollmann},\ and\ \citenamefont {von
  Keyserlingk}}]{rakovszky2018diffusive}%
  \BibitemOpen
  \bibfield  {author} {\bibinfo {author} {\bibfnamefont {T.}~\bibnamefont
  {Rakovszky}}, \bibinfo {author} {\bibfnamefont {F.}~\bibnamefont
  {Pollmann}},\ and\ \bibinfo {author} {\bibfnamefont {C.~W.}\ \bibnamefont
  {von Keyserlingk}},\ }\bibfield  {title} {\bibinfo {title} {Diffusive
  hydrodynamics of out-of-time-ordered correlators with charge conservation},\
  }\href {https://doi.org/10.1103/PhysRevX.8.031058} {\bibfield  {journal}
  {\bibinfo  {journal} {Phys. Rev. X}\ }\textbf {\bibinfo {volume} {8}},\
  \bibinfo {pages} {031058} (\bibinfo {year} {2018})}\BibitemShut {NoStop}%
\bibitem [{\citenamefont {Khemani}\ \emph {et~al.}(2018)\citenamefont
  {Khemani}, \citenamefont {Huse},\ and\ \citenamefont
  {Nahum}}]{khemani2018velocity}%
  \BibitemOpen
  \bibfield  {author} {\bibinfo {author} {\bibfnamefont {V.}~\bibnamefont
  {Khemani}}, \bibinfo {author} {\bibfnamefont {D.~A.}\ \bibnamefont {Huse}},\
  and\ \bibinfo {author} {\bibfnamefont {A.}~\bibnamefont {Nahum}},\ }\bibfield
   {title} {\bibinfo {title} {Velocity-dependent lyapunov exponents in
  many-body quantum, semiclassical, and classical chaos},\ }\href
  {https://doi.org/10.1103/PhysRevB.98.144304} {\bibfield  {journal} {\bibinfo
  {journal} {Phys. Rev. B}\ }\textbf {\bibinfo {volume} {98}},\ \bibinfo
  {pages} {144304} (\bibinfo {year} {2018})}\BibitemShut {NoStop}%
\bibitem [{\citenamefont {Das}\ \emph {et~al.}(2018)\citenamefont {Das},
  \citenamefont {Chakrabarty}, \citenamefont {Dhar}, \citenamefont {Kundu},
  \citenamefont {Huse}, \citenamefont {Moessner}, \citenamefont {Ray},\ and\
  \citenamefont {Bhattacharjee}}]{das2018light}%
  \BibitemOpen
  \bibfield  {author} {\bibinfo {author} {\bibfnamefont {A.}~\bibnamefont
  {Das}}, \bibinfo {author} {\bibfnamefont {S.}~\bibnamefont {Chakrabarty}},
  \bibinfo {author} {\bibfnamefont {A.}~\bibnamefont {Dhar}}, \bibinfo {author}
  {\bibfnamefont {A.}~\bibnamefont {Kundu}}, \bibinfo {author} {\bibfnamefont
  {D.~A.}\ \bibnamefont {Huse}}, \bibinfo {author} {\bibfnamefont
  {R.}~\bibnamefont {Moessner}}, \bibinfo {author} {\bibfnamefont {S.~S.}\
  \bibnamefont {Ray}},\ and\ \bibinfo {author} {\bibfnamefont {S.}~\bibnamefont
  {Bhattacharjee}},\ }\bibfield  {title} {\bibinfo {title} {Light-cone
  spreading of perturbations and the butterfly effect in a classical spin
  chain},\ }\href {https://doi.org/10.1103/PhysRevLett.121.024101} {\bibfield
  {journal} {\bibinfo  {journal} {Phys. Rev. Lett.}\ }\textbf {\bibinfo
  {volume} {121}},\ \bibinfo {pages} {024101} (\bibinfo {year}
  {2018})}\BibitemShut {NoStop}%
\bibitem [{\citenamefont {Bilitewski}\ \emph {et~al.}(2018)\citenamefont
  {Bilitewski}, \citenamefont {Bhattacharjee},\ and\ \citenamefont
  {Moessner}}]{bilitewski2018temperature}%
  \BibitemOpen
  \bibfield  {author} {\bibinfo {author} {\bibfnamefont {T.}~\bibnamefont
  {Bilitewski}}, \bibinfo {author} {\bibfnamefont {S.}~\bibnamefont
  {Bhattacharjee}},\ and\ \bibinfo {author} {\bibfnamefont {R.}~\bibnamefont
  {Moessner}},\ }\bibfield  {title} {\bibinfo {title} {Temperature dependence
  of the butterfly effect in a classical many-body system},\ }\href
  {https://doi.org/10.1103/PhysRevLett.121.250602} {\bibfield  {journal}
  {\bibinfo  {journal} {Phys. Rev. Lett.}\ }\textbf {\bibinfo {volume} {121}},\
  \bibinfo {pages} {250602} (\bibinfo {year} {2018})}\BibitemShut {NoStop}%
\bibitem [{\citenamefont {Murugan}\ \emph {et~al.}(2021)\citenamefont
  {Murugan}, \citenamefont {Kumar}, \citenamefont {Bhattacharjee},\ and\
  \citenamefont {Ray}}]{murugan2021many}%
  \BibitemOpen
  \bibfield  {author} {\bibinfo {author} {\bibfnamefont {S.~D.}\ \bibnamefont
  {Murugan}}, \bibinfo {author} {\bibfnamefont {D.}~\bibnamefont {Kumar}},
  \bibinfo {author} {\bibfnamefont {S.}~\bibnamefont {Bhattacharjee}},\ and\
  \bibinfo {author} {\bibfnamefont {S.~S.}\ \bibnamefont {Ray}},\ }\bibfield
  {title} {\bibinfo {title} {Many-body chaos in thermalized fluids},\ }\href
  {https://doi.org/10.1103/PhysRevLett.127.124501} {\bibfield  {journal}
  {\bibinfo  {journal} {Phys. Rev. Lett.}\ }\textbf {\bibinfo {volume} {127}},\
  \bibinfo {pages} {124501} (\bibinfo {year} {2021})}\BibitemShut {NoStop}%
\bibitem [{\citenamefont {Liu}\ \emph {et~al.}(2021)\citenamefont {Liu},
  \citenamefont {Willsher}, \citenamefont {Bilitewski}, \citenamefont {Li},
  \citenamefont {Smith}, \citenamefont {Christensen}, \citenamefont
  {Moessner},\ and\ \citenamefont {Knolle}}]{liu2021butterfly}%
  \BibitemOpen
  \bibfield  {author} {\bibinfo {author} {\bibfnamefont {S.-W.}\ \bibnamefont
  {Liu}}, \bibinfo {author} {\bibfnamefont {J.}~\bibnamefont {Willsher}},
  \bibinfo {author} {\bibfnamefont {T.}~\bibnamefont {Bilitewski}}, \bibinfo
  {author} {\bibfnamefont {J.-J.}\ \bibnamefont {Li}}, \bibinfo {author}
  {\bibfnamefont {A.}~\bibnamefont {Smith}}, \bibinfo {author} {\bibfnamefont
  {K.}~\bibnamefont {Christensen}}, \bibinfo {author} {\bibfnamefont
  {R.}~\bibnamefont {Moessner}},\ and\ \bibinfo {author} {\bibfnamefont
  {J.}~\bibnamefont {Knolle}},\ }\bibfield  {title} {\bibinfo {title}
  {Butterfly effect and spatial structure of information spreading in a chaotic
  cellular automaton},\ }\href {https://doi.org/10.1103/PhysRevB.103.094109}
  {\bibfield  {journal} {\bibinfo  {journal} {Phys. Rev. B}\ }\textbf {\bibinfo
  {volume} {103}},\ \bibinfo {pages} {094109} (\bibinfo {year}
  {2021})}\BibitemShut {NoStop}%
\bibitem [{\citenamefont {Bilitewski}\ \emph {et~al.}(2021)\citenamefont
  {Bilitewski}, \citenamefont {Bhattacharjee},\ and\ \citenamefont
  {Moessner}}]{bilitewski2021classical}%
  \BibitemOpen
  \bibfield  {author} {\bibinfo {author} {\bibfnamefont {T.}~\bibnamefont
  {Bilitewski}}, \bibinfo {author} {\bibfnamefont {S.}~\bibnamefont
  {Bhattacharjee}},\ and\ \bibinfo {author} {\bibfnamefont {R.}~\bibnamefont
  {Moessner}},\ }\bibfield  {title} {\bibinfo {title} {Classical many-body
  chaos with and without quasiparticles},\ }\href
  {https://doi.org/10.1103/PhysRevB.103.174302} {\bibfield  {journal} {\bibinfo
   {journal} {Phys. Rev. B}\ }\textbf {\bibinfo {volume} {103}},\ \bibinfo
  {pages} {174302} (\bibinfo {year} {2021})}\BibitemShut {NoStop}%
\bibitem [{\citenamefont {Gibbs}\ and\ \citenamefont
  {DiMarzio}(1958)}]{gibbs1958nature}%
  \BibitemOpen
  \bibfield  {author} {\bibinfo {author} {\bibfnamefont {J.~H.}\ \bibnamefont
  {Gibbs}}\ and\ \bibinfo {author} {\bibfnamefont {E.~A.}\ \bibnamefont
  {DiMarzio}},\ }\bibfield  {title} {\bibinfo {title} {Nature of the glass
  transition and the glassy state},\ }\href {https://doi.org/10.1063/1.1744141}
  {\bibfield  {journal} {\bibinfo  {journal} {The Journal of Chemical Physics}\
  }\textbf {\bibinfo {volume} {28}},\ \bibinfo {pages} {373} (\bibinfo {year}
  {1958})}\BibitemShut {NoStop}%
\bibitem [{\citenamefont {J\"ackle}(1986)}]{jackle1986models}%
  \BibitemOpen
  \bibfield  {author} {\bibinfo {author} {\bibfnamefont {J.}~\bibnamefont
  {J\"ackle}},\ }\bibfield  {title} {\bibinfo {title} {Models of the glass
  transition},\ }\href {https://doi.org/10.1088/0034-4885/49/2/002} {\bibfield
  {journal} {\bibinfo  {journal} {Reports on Progress in Physics}\ }\textbf
  {\bibinfo {volume} {49}},\ \bibinfo {pages} {171} (\bibinfo {year}
  {1986})}\BibitemShut {NoStop}%
\bibitem [{\citenamefont {M{\'{e}}zard}\ and\ \citenamefont
  {Parisi}(2000)}]{mezard2000statistical}%
  \BibitemOpen
  \bibfield  {author} {\bibinfo {author} {\bibfnamefont {M.}~\bibnamefont
  {M{\'{e}}zard}}\ and\ \bibinfo {author} {\bibfnamefont {G.}~\bibnamefont
  {Parisi}},\ }\bibfield  {title} {\bibinfo {title} {Statistical physics of
  structural glasses},\ }\href {https://doi.org/10.1088/0953-8984/12/29/336}
  {\bibfield  {journal} {\bibinfo  {journal} {Journal of Physics: Condensed
  Matter}\ }\textbf {\bibinfo {volume} {12}},\ \bibinfo {pages} {6655}
  (\bibinfo {year} {2000})}\BibitemShut {NoStop}%
\bibitem [{\citenamefont {Edwards}\ and\ \citenamefont
  {Anderson}(1975)}]{edwards1975theory}%
  \BibitemOpen
  \bibfield  {author} {\bibinfo {author} {\bibfnamefont {S.~F.}\ \bibnamefont
  {Edwards}}\ and\ \bibinfo {author} {\bibfnamefont {P.~W.}\ \bibnamefont
  {Anderson}},\ }\bibfield  {title} {\bibinfo {title} {Theory of spin
  glasses},\ }\href {https://doi.org/10.1088/0305-4608/5/5/017} {\bibfield
  {journal} {\bibinfo  {journal} {Journal of Physics F: Metal Physics}\
  }\textbf {\bibinfo {volume} {5}},\ \bibinfo {pages} {965} (\bibinfo {year}
  {1975})}\BibitemShut {NoStop}%
\bibitem [{\citenamefont {Sherrington}\ and\ \citenamefont
  {Kirkpatrick}(1975)}]{sherrington1975solvable}%
  \BibitemOpen
  \bibfield  {author} {\bibinfo {author} {\bibfnamefont {D.}~\bibnamefont
  {Sherrington}}\ and\ \bibinfo {author} {\bibfnamefont {S.}~\bibnamefont
  {Kirkpatrick}},\ }\bibfield  {title} {\bibinfo {title} {Solvable model of a
  spin-glass},\ }\href {https://doi.org/10.1103/PhysRevLett.35.1792} {\bibfield
   {journal} {\bibinfo  {journal} {Phys. Rev. Lett.}\ }\textbf {\bibinfo
  {volume} {35}},\ \bibinfo {pages} {1792} (\bibinfo {year}
  {1975})}\BibitemShut {NoStop}%
\bibitem [{\citenamefont {Binder}\ and\ \citenamefont
  {Young}(1986)}]{binder1986spin}%
  \BibitemOpen
  \bibfield  {author} {\bibinfo {author} {\bibfnamefont {K.}~\bibnamefont
  {Binder}}\ and\ \bibinfo {author} {\bibfnamefont {A.~P.}\ \bibnamefont
  {Young}},\ }\bibfield  {title} {\bibinfo {title} {Spin glasses: Experimental
  facts, theoretical concepts, and open questions},\ }\href
  {https://doi.org/10.1103/RevModPhys.58.801} {\bibfield  {journal} {\bibinfo
  {journal} {Rev. Mod. Phys.}\ }\textbf {\bibinfo {volume} {58}},\ \bibinfo
  {pages} {801} (\bibinfo {year} {1986})}\BibitemShut {NoStop}%
\bibitem [{\citenamefont {Mezard}\ \emph {et~al.}(1986)\citenamefont {Mezard},
  \citenamefont {Parisi},\ and\ \citenamefont {Virasoro}}]{mezard1986spin}%
  \BibitemOpen
  \bibfield  {author} {\bibinfo {author} {\bibfnamefont {M.}~\bibnamefont
  {Mezard}}, \bibinfo {author} {\bibfnamefont {G.}~\bibnamefont {Parisi}},\
  and\ \bibinfo {author} {\bibfnamefont {M.}~\bibnamefont {Virasoro}},\ }\href
  {https://doi.org/10.1142/0271} {\emph {\bibinfo {title} {Spin Glass Theory
  and Beyond}}}\ (\bibinfo  {publisher} {World Scientific},\ \bibinfo {address}
  {Singapore},\ \bibinfo {year} {1986})\BibitemShut {NoStop}%
\bibitem [{\citenamefont {Anderson}(1958)}]{anderson1958absence}%
  \BibitemOpen
  \bibfield  {author} {\bibinfo {author} {\bibfnamefont {P.~W.}\ \bibnamefont
  {Anderson}},\ }\bibfield  {title} {\bibinfo {title} {Absence of diffusion in
  certain random lattices},\ }\href {https://doi.org/10.1103/PhysRev.109.1492}
  {\bibfield  {journal} {\bibinfo  {journal} {Phys. Rev.}\ }\textbf {\bibinfo
  {volume} {109}},\ \bibinfo {pages} {1492} (\bibinfo {year}
  {1958})}\BibitemShut {NoStop}%
\bibitem [{\citenamefont {Lee}\ and\ \citenamefont
  {Ramakrishnan}(1985)}]{lee1985disordered}%
  \BibitemOpen
  \bibfield  {author} {\bibinfo {author} {\bibfnamefont {P.~A.}\ \bibnamefont
  {Lee}}\ and\ \bibinfo {author} {\bibfnamefont {T.~V.}\ \bibnamefont
  {Ramakrishnan}},\ }\bibfield  {title} {\bibinfo {title} {Disordered
  electronic systems},\ }\href {https://doi.org/10.1103/RevModPhys.57.287}
  {\bibfield  {journal} {\bibinfo  {journal} {Rev. Mod. Phys.}\ }\textbf
  {\bibinfo {volume} {57}},\ \bibinfo {pages} {287} (\bibinfo {year}
  {1985})}\BibitemShut {NoStop}%
\bibitem [{\citenamefont {Nandkishore}\ and\ \citenamefont
  {Huse}(2015)}]{nandkishore2015many}%
  \BibitemOpen
  \bibfield  {author} {\bibinfo {author} {\bibfnamefont {R.}~\bibnamefont
  {Nandkishore}}\ and\ \bibinfo {author} {\bibfnamefont {D.~A.}\ \bibnamefont
  {Huse}},\ }\bibfield  {title} {\bibinfo {title} {Many-body localization and
  thermalization in quantum statistical mechanics},\ }\href
  {https://doi.org/10.1146/annurev-conmatphys-031214-014726} {\bibfield
  {journal} {\bibinfo  {journal} {Annu. Rev. Condens. Matter Phys.}\ }\textbf
  {\bibinfo {volume} {6}},\ \bibinfo {pages} {15} (\bibinfo {year}
  {2015})}\BibitemShut {NoStop}%
\bibitem [{\citenamefont {Alet}\ and\ \citenamefont
  {Laflorencie}(2018)}]{alet2018many}%
  \BibitemOpen
  \bibfield  {author} {\bibinfo {author} {\bibfnamefont {F.}~\bibnamefont
  {Alet}}\ and\ \bibinfo {author} {\bibfnamefont {N.}~\bibnamefont
  {Laflorencie}},\ }\bibfield  {title} {\bibinfo {title} {Many-body
  localization: an introduction and selected topics},\ }\href
  {https://doi.org/https://doi.org/10.1016/j.crhy.2018.03.003} {\bibfield
  {journal} {\bibinfo  {journal} {Comptes Rendus Physique}\ }\textbf {\bibinfo
  {volume} {19}},\ \bibinfo {pages} {498} (\bibinfo {year} {2018})}\BibitemShut
  {NoStop}%
\bibitem [{\citenamefont {Abanin}\ \emph {et~al.}(2019)\citenamefont {Abanin},
  \citenamefont {Altman}, \citenamefont {Bloch},\ and\ \citenamefont
  {Serbyn}}]{abanin2019colloquium}%
  \BibitemOpen
  \bibfield  {author} {\bibinfo {author} {\bibfnamefont {D.~A.}\ \bibnamefont
  {Abanin}}, \bibinfo {author} {\bibfnamefont {E.}~\bibnamefont {Altman}},
  \bibinfo {author} {\bibfnamefont {I.}~\bibnamefont {Bloch}},\ and\ \bibinfo
  {author} {\bibfnamefont {M.}~\bibnamefont {Serbyn}},\ }\bibfield  {title}
  {\bibinfo {title} {Colloquium: Many-body localization, thermalization, and
  entanglement},\ }\href {https://doi.org/10.1103/RevModPhys.91.021001}
  {\bibfield  {journal} {\bibinfo  {journal} {Rev. Mod. Phys.}\ }\textbf
  {\bibinfo {volume} {91}},\ \bibinfo {pages} {021001} (\bibinfo {year}
  {2019})}\BibitemShut {NoStop}%
\bibitem [{\citenamefont {Fredrickson}\ and\ \citenamefont
  {Andersen}(1984)}]{fredrickson1984kinetic}%
  \BibitemOpen
  \bibfield  {author} {\bibinfo {author} {\bibfnamefont {G.~H.}\ \bibnamefont
  {Fredrickson}}\ and\ \bibinfo {author} {\bibfnamefont {H.~C.}\ \bibnamefont
  {Andersen}},\ }\bibfield  {title} {\bibinfo {title} {Kinetic {I}sing model of
  the {G}lass {T}ransition},\ }\href
  {https://doi.org/10.1103/PhysRevLett.53.1244} {\bibfield  {journal} {\bibinfo
   {journal} {Phys. Rev. Lett.}\ }\textbf {\bibinfo {volume} {53}},\ \bibinfo
  {pages} {1244} (\bibinfo {year} {1984})}\BibitemShut {NoStop}%
\bibitem [{\citenamefont {Fredrickson}\ and\ \citenamefont
  {Andersen}(1985)}]{fredrickson1985facilitated}%
  \BibitemOpen
  \bibfield  {author} {\bibinfo {author} {\bibfnamefont {G.~H.}\ \bibnamefont
  {Fredrickson}}\ and\ \bibinfo {author} {\bibfnamefont {H.~C.}\ \bibnamefont
  {Andersen}},\ }\bibfield  {title} {\bibinfo {title} {Facilitated kinetic
  {I}sing models and the glass transition},\ }\href
  {https://aip.scitation.org/doi/10.1063/1.449662} {\bibfield  {journal}
  {\bibinfo  {journal} {J. Chem. Phys.}\ }\textbf {\bibinfo {volume} {83}},\
  \bibinfo {pages} {5822} (\bibinfo {year} {1985})}\BibitemShut {NoStop}%
\bibitem [{\citenamefont {J{\"a}ckle}\ and\ \citenamefont
  {Eisinger}(1991)}]{jackle1991hierarchically}%
  \BibitemOpen
  \bibfield  {author} {\bibinfo {author} {\bibfnamefont {J.}~\bibnamefont
  {J{\"a}ckle}}\ and\ \bibinfo {author} {\bibfnamefont {S.}~\bibnamefont
  {Eisinger}},\ }\bibfield  {title} {\bibinfo {title} {A hierarchically
  constrained kinetic ising model},\ }\href
  {https://doi.org/https://doi.org/10.1007/BF01453764} {\bibfield  {journal}
  {\bibinfo  {journal} {Z. Phys. B Condens. Matter}\ }\textbf {\bibinfo
  {volume} {84}},\ \bibinfo {pages} {115} (\bibinfo {year} {1991})}\BibitemShut
  {NoStop}%
\bibitem [{\citenamefont {Sollich}\ and\ \citenamefont
  {Evans}(1999)}]{sollich1999glassy}%
  \BibitemOpen
  \bibfield  {author} {\bibinfo {author} {\bibfnamefont {P.}~\bibnamefont
  {Sollich}}\ and\ \bibinfo {author} {\bibfnamefont {M.~R.}\ \bibnamefont
  {Evans}},\ }\bibfield  {title} {\bibinfo {title} {Glassy time-scale
  divergence and anomalous coarsening in a kinetically constrained spin
  chain},\ }\href {https://doi.org/10.1103/PhysRevLett.83.3238} {\bibfield
  {journal} {\bibinfo  {journal} {Phys. Rev. Lett.}\ }\textbf {\bibinfo
  {volume} {83}},\ \bibinfo {pages} {3238} (\bibinfo {year}
  {1999})}\BibitemShut {NoStop}%
\bibitem [{\citenamefont {Sollich}\ and\ \citenamefont
  {Evans}(2003)}]{sollich2003glassy}%
  \BibitemOpen
  \bibfield  {author} {\bibinfo {author} {\bibfnamefont {P.}~\bibnamefont
  {Sollich}}\ and\ \bibinfo {author} {\bibfnamefont {M.~R.}\ \bibnamefont
  {Evans}},\ }\bibfield  {title} {\bibinfo {title} {Glassy dynamics in the
  asymmetrically constrained kinetic ising chain},\ }\href
  {https://doi.org/10.1103/PhysRevE.68.031504} {\bibfield  {journal} {\bibinfo
  {journal} {Phys. Rev. E}\ }\textbf {\bibinfo {volume} {68}},\ \bibinfo
  {pages} {031504} (\bibinfo {year} {2003})}\BibitemShut {NoStop}%
\bibitem [{\citenamefont {Ritort}\ and\ \citenamefont
  {Sollich}(2003)}]{ritort2003glassy}%
  \BibitemOpen
  \bibfield  {author} {\bibinfo {author} {\bibfnamefont {F.}~\bibnamefont
  {Ritort}}\ and\ \bibinfo {author} {\bibfnamefont {P.}~\bibnamefont
  {Sollich}},\ }\bibfield  {title} {\bibinfo {title} {Glassy dynamics of
  kinetically constrained models},\ }\href
  {https://www.tandfonline.com/doi/abs/10.1080/0001873031000093582} {\bibfield
  {journal} {\bibinfo  {journal} {Advances in Physics}\ }\textbf {\bibinfo
  {volume} {52}},\ \bibinfo {pages} {219} (\bibinfo {year} {2003})}\BibitemShut
  {NoStop}%
\bibitem [{\citenamefont {Garrahan}\ \emph {et~al.}(2011)\citenamefont
  {Garrahan}, \citenamefont {Sollich},\ and\ \citenamefont
  {Toninelli}}]{garrahan2011kinetically}%
  \BibitemOpen
  \bibfield  {author} {\bibinfo {author} {\bibfnamefont {J.~P.}\ \bibnamefont
  {Garrahan}}, \bibinfo {author} {\bibfnamefont {P.}~\bibnamefont {Sollich}},\
  and\ \bibinfo {author} {\bibfnamefont {C.}~\bibnamefont {Toninelli}},\
  }\bibfield  {title} {\bibinfo {title} {Kinetically constrained models},\ }in\
  \href
  {http://www.oxfordscholarship.com/view/10.1093/acprof:oso/9780199691470.001.0001/acprof-9780199691470-chapter-10}
  {\emph {\bibinfo {booktitle} {Dynamical heterogeneities in glasses, colloids,
  and granular media}}},\ \bibinfo {editor} {edited by\ \bibinfo {editor}
  {\bibfnamefont {L.}~\bibnamefont {Berthier}}, \bibinfo {editor}
  {\bibfnamefont {G.}~\bibnamefont {Biroli}}, \bibinfo {editor} {\bibfnamefont
  {J.-P.}\ \bibnamefont {Bouchaud}}, \bibinfo {editor} {\bibfnamefont
  {L.}~\bibnamefont {Cipelletti}},\ and\ \bibinfo {editor} {\bibfnamefont
  {W.}~\bibnamefont {van Saarloos}}}\ (\bibinfo  {publisher} {Oxford University
  Press},\ \bibinfo {address} {Oxford},\ \bibinfo {year} {2011})\BibitemShut
  {NoStop}%
\bibitem [{\citenamefont {van Horssen}\ \emph {et~al.}(2015)\citenamefont {van
  Horssen}, \citenamefont {Levi},\ and\ \citenamefont
  {Garrahan}}]{horssen2015dynamics}%
  \BibitemOpen
  \bibfield  {author} {\bibinfo {author} {\bibfnamefont {M.}~\bibnamefont {van
  Horssen}}, \bibinfo {author} {\bibfnamefont {E.}~\bibnamefont {Levi}},\ and\
  \bibinfo {author} {\bibfnamefont {J.~P.}\ \bibnamefont {Garrahan}},\
  }\bibfield  {title} {\bibinfo {title} {Dynamics of many-body localization in
  a translation-invariant quantum glass model},\ }\href
  {https://doi.org/10.1103/PhysRevB.92.100305} {\bibfield  {journal} {\bibinfo
  {journal} {Phys. Rev. B}\ }\textbf {\bibinfo {volume} {92}},\ \bibinfo
  {pages} {100305} (\bibinfo {year} {2015})}\BibitemShut {NoStop}%
\bibitem [{\citenamefont {Lan}\ \emph {et~al.}(2018)\citenamefont {Lan},
  \citenamefont {van Horssen}, \citenamefont {Powell},\ and\ \citenamefont
  {Garrahan}}]{lan2018quantum}%
  \BibitemOpen
  \bibfield  {author} {\bibinfo {author} {\bibfnamefont {Z.}~\bibnamefont
  {Lan}}, \bibinfo {author} {\bibfnamefont {M.}~\bibnamefont {van Horssen}},
  \bibinfo {author} {\bibfnamefont {S.}~\bibnamefont {Powell}},\ and\ \bibinfo
  {author} {\bibfnamefont {J.~P.}\ \bibnamefont {Garrahan}},\ }\bibfield
  {title} {\bibinfo {title} {Quantum slow relaxation and metastability due to
  dynamical constraints},\ }\href
  {https://doi.org/10.1103/PhysRevLett.121.040603} {\bibfield  {journal}
  {\bibinfo  {journal} {Phys. Rev. Lett.}\ }\textbf {\bibinfo {volume} {121}},\
  \bibinfo {pages} {040603} (\bibinfo {year} {2018})}\BibitemShut {NoStop}%
\bibitem [{\citenamefont {Pancotti}\ \emph {et~al.}(2020)\citenamefont
  {Pancotti}, \citenamefont {Giudice}, \citenamefont {Cirac}, \citenamefont
  {Garrahan},\ and\ \citenamefont {Ba\~nuls}}]{pancotti2020quantum}%
  \BibitemOpen
  \bibfield  {author} {\bibinfo {author} {\bibfnamefont {N.}~\bibnamefont
  {Pancotti}}, \bibinfo {author} {\bibfnamefont {G.}~\bibnamefont {Giudice}},
  \bibinfo {author} {\bibfnamefont {J.~I.}\ \bibnamefont {Cirac}}, \bibinfo
  {author} {\bibfnamefont {J.~P.}\ \bibnamefont {Garrahan}},\ and\ \bibinfo
  {author} {\bibfnamefont {M.~C.}\ \bibnamefont {Ba\~nuls}},\ }\bibfield
  {title} {\bibinfo {title} {Quantum east model: Localization, nonthermal
  eigenstates, and slow dynamics},\ }\href
  {https://doi.org/10.1103/PhysRevX.10.021051} {\bibfield  {journal} {\bibinfo
  {journal} {Phys. Rev. X}\ }\textbf {\bibinfo {volume} {10}},\ \bibinfo
  {pages} {021051} (\bibinfo {year} {2020})}\BibitemShut {NoStop}%
\bibitem [{\citenamefont {Roy}\ and\ \citenamefont
  {Lazarides}(2020)}]{roy2020strong}%
  \BibitemOpen
  \bibfield  {author} {\bibinfo {author} {\bibfnamefont {S.}~\bibnamefont
  {Roy}}\ and\ \bibinfo {author} {\bibfnamefont {A.}~\bibnamefont
  {Lazarides}},\ }\bibfield  {title} {\bibinfo {title} {Strong ergodicity
  breaking due to local constraints in a quantum system},\ }\href
  {https://doi.org/10.1103/PhysRevResearch.2.023159} {\bibfield  {journal}
  {\bibinfo  {journal} {Phys. Rev. Research}\ }\textbf {\bibinfo {volume}
  {2}},\ \bibinfo {pages} {023159} (\bibinfo {year} {2020})}\BibitemShut
  {NoStop}%
\bibitem [{\citenamefont {Pai}\ \emph {et~al.}(2019)\citenamefont {Pai},
  \citenamefont {Pretko},\ and\ \citenamefont
  {Nandkishore}}]{pai2019localisation}%
  \BibitemOpen
  \bibfield  {author} {\bibinfo {author} {\bibfnamefont {S.}~\bibnamefont
  {Pai}}, \bibinfo {author} {\bibfnamefont {M.}~\bibnamefont {Pretko}},\ and\
  \bibinfo {author} {\bibfnamefont {R.~M.}\ \bibnamefont {Nandkishore}},\
  }\bibfield  {title} {\bibinfo {title} {Localization in fractonic random
  circuits},\ }\href {https://doi.org/10.1103/PhysRevX.9.021003} {\bibfield
  {journal} {\bibinfo  {journal} {Phys. Rev. X}\ }\textbf {\bibinfo {volume}
  {9}},\ \bibinfo {pages} {021003} (\bibinfo {year} {2019})}\BibitemShut
  {NoStop}%
\bibitem [{\citenamefont {Khemani}\ \emph {et~al.}(2020)\citenamefont
  {Khemani}, \citenamefont {Hermele},\ and\ \citenamefont
  {Nandkishore}}]{khemani2020localisation}%
  \BibitemOpen
  \bibfield  {author} {\bibinfo {author} {\bibfnamefont {V.}~\bibnamefont
  {Khemani}}, \bibinfo {author} {\bibfnamefont {M.}~\bibnamefont {Hermele}},\
  and\ \bibinfo {author} {\bibfnamefont {R.}~\bibnamefont {Nandkishore}},\
  }\bibfield  {title} {\bibinfo {title} {Localization from hilbert space
  shattering: {F}rom theory to physical realizations},\ }\href
  {https://doi.org/10.1103/PhysRevB.101.174204} {\bibfield  {journal} {\bibinfo
   {journal} {Phys. Rev. B}\ }\textbf {\bibinfo {volume} {101}},\ \bibinfo
  {pages} {174204} (\bibinfo {year} {2020})}\BibitemShut {NoStop}%
\bibitem [{\citenamefont {Sala}\ \emph {et~al.}(2020)\citenamefont {Sala},
  \citenamefont {Rakovszky}, \citenamefont {Verresen}, \citenamefont {Knap},\
  and\ \citenamefont {Pollmann}}]{sala2020ergodicity}%
  \BibitemOpen
  \bibfield  {author} {\bibinfo {author} {\bibfnamefont {P.}~\bibnamefont
  {Sala}}, \bibinfo {author} {\bibfnamefont {T.}~\bibnamefont {Rakovszky}},
  \bibinfo {author} {\bibfnamefont {R.}~\bibnamefont {Verresen}}, \bibinfo
  {author} {\bibfnamefont {M.}~\bibnamefont {Knap}},\ and\ \bibinfo {author}
  {\bibfnamefont {F.}~\bibnamefont {Pollmann}},\ }\bibfield  {title} {\bibinfo
  {title} {Ergodicity breaking arising from {H}ilbert space fragmentation in
  dipole-conserving hamiltonians},\ }\href
  {https://doi.org/10.1103/PhysRevX.10.011047} {\bibfield  {journal} {\bibinfo
  {journal} {Phys. Rev. X}\ }\textbf {\bibinfo {volume} {10}},\ \bibinfo
  {pages} {011047} (\bibinfo {year} {2020})}\BibitemShut {NoStop}%
\bibitem [{\citenamefont {Feldmeier}\ and\ \citenamefont
  {Knap}(2021)}]{feldmeir2021critically}%
  \BibitemOpen
  \bibfield  {author} {\bibinfo {author} {\bibfnamefont {J.}~\bibnamefont
  {Feldmeier}}\ and\ \bibinfo {author} {\bibfnamefont {M.}~\bibnamefont
  {Knap}},\ }\bibfield  {title} {\bibinfo {title} {Critically slow operator
  dynamics in constrained many-body systems},\ }\href
  {https://doi.org/10.1103/PhysRevLett.127.235301} {\bibfield  {journal}
  {\bibinfo  {journal} {Phys. Rev. Lett.}\ }\textbf {\bibinfo {volume} {127}},\
  \bibinfo {pages} {235301} (\bibinfo {year} {2021})}\BibitemShut {NoStop}%
\bibitem [{Note1()}]{Note1}%
  \BibitemOpen
  \bibinfo {note} {As shown in Ref.~\cite {das2018light}, on canonical
  quantisation of the Poisson brackets $\mathinner {\langle {D(x,t)}\rangle
  }\to -(\varepsilon ^2/\hbar ^2)\protect \mathrm {Tr}[[\protect \hat {\protect
  \mathbf {S}}_x(t),\protect \hat {\protect \mathbf {z}}\cdot \protect \hat
  {\protect \mathbf {S}}_0(0)]^2]$ which is nothing but the quantum
  OTOC.}\BibitemShut {Stop}%
\bibitem [{Note2()}]{Note2}%
  \BibitemOpen
  \bibinfo {note} {Strictly speaking, this divergence rules out the ballistic
  spreading of the front. Subsequent results on melting of an initially frozen
  island (Fig.~\ref {fig:melting}) provides evidence for the complete arrest of
  the lightcone.}\BibitemShut {Stop}%
\bibitem [{Note3()}]{Note3}%
  \BibitemOpen
  \bibinfo {note} {A systematic study of the critical behaviour incorporating
  finite-size and finite-time effects is beyond the scope and interest of this
  work.}\BibitemShut {Stop}%
\bibitem [{sup()}]{supp}%
  \BibitemOpen
  \href@noop {} {}\bibinfo {note} {See supplementary material at
  [URL].}\BibitemShut {Stop}%
\bibitem [{\citenamefont {Deger}\ \emph {et~al.}(2022)\citenamefont {Deger},
  \citenamefont {Lazarides},\ and\ \citenamefont {Roy}}]{deger2022phase}%
  \BibitemOpen
  \bibfield  {author} {\bibinfo {author} {\bibfnamefont {A.}~\bibnamefont
  {Deger}}, \bibinfo {author} {\bibfnamefont {A.}~\bibnamefont {Lazarides}},\
  and\ \bibinfo {author} {\bibfnamefont {S.}~\bibnamefont {Roy}},\ }\href
  {https://doi.org/10.48550/ARXIV.2206.07724} {\bibinfo {title} {Constrained
  dynamics and directed percolation}} (\bibinfo {year} {2022})\BibitemShut
  {NoStop}%
\bibitem [{\citenamefont {Preskill}(2018)}]{preskill2018quantum}%
  \BibitemOpen
  \bibfield  {author} {\bibinfo {author} {\bibfnamefont {J.}~\bibnamefont
  {Preskill}},\ }\bibfield  {title} {\bibinfo {title} {Quantum computing in the
  {NISQ} era and beyond},\ }\href {https://doi.org/10.22331/q-2018-08-06-79}
  {\bibfield  {journal} {\bibinfo  {journal} {Quantum}\ }\textbf {\bibinfo
  {volume} {2}},\ \bibinfo {pages} {79} (\bibinfo {year} {2018})}\BibitemShut
  {NoStop}%
\end{thebibliography}%

\clearpage
\onecolumngrid

\setcounter{equation}{0}
\setcounter{figure}{0}
\def\theequation{S\arabic{equation}}
\def\thefigure{S\arabic{figure}}
\begin{center}
{\textbf{{Supplementary Material: Arresting classical many-body chaos by kinetic constraints}}}\\
Aydin Deger, Sthitadhi Roy, and Achilleas Lazarides
\end{center}
\bigskip

\begin{quote}
This supplementary material has two sections. In Sec.~\ref{sec:ham}, we present results for a time-independent Hamiltonian model. In Sec.~\ref{sec:2d}, we present results for a Floquet model similar to that in the main text but defined on a square lattice in two spatial dimensions. In both of these cases, we again see clear evidence of a dynamical phase transition between a phase where the classical OTOC has a ballistic lightcone and another where it is completely arrested. This shows the generality of our results.
\end{quote}

\section{Results for time-independent Hamiltonians \label{sec:ham}}
We consider a time-independent Hamiltonian model which is the average of the two Hamiltonians in the Floquet model. It is given by,
\begin{align}
    H_\mathrm{Ising} = \sum_{x}[S_x^zS_{x+1}^z + h S^z_x + g S^x_x]\,.
    \label{eq:Hising}
\end{align}

The Hamiltonian equations of motion, which automatically conserve total energy are given by 
\begin{equation}
    \partial_t\mathbf{S}_x = \Theta_x\{\mathbf{S}_x,H\}\,,
\end{equation}
where $\Theta_x$ encodes the kinetic constraints as before (see Eq.~3 of the main text), 
\begin{align}
\{f,g\} = \sum_x\sum_{\alpha,\beta,\gamma}\epsilon^{\alpha\beta\gamma}S^\gamma_x (\partial f/\partial S^\alpha_x)(\partial g/\partial S^\beta_x)
\end{align}
is the spin Poisson bracket and $\epsilon$ is the Levi-Civita tensor. For $H_\mathrm{Ising}$ in Eq.~\ref{eq:Hising}, the equations are given explicitly by
\begin{align}
\partial_t S_x^x &=\Theta_x [-S_x^y(S^z_{x-1}+S^z_{x+1}+h)]\,,\nonumber\\
\partial_t S_x^y &=\Theta_x [S_x^x(S^z_{x-1}+S^z_{x+1}+h)-gS^z_x]\,,\label{eq:eomising}\\
\partial_t S_x^z &=\Theta_x gS_x^y\nonumber\,.
\end{align}

\begin{figure}[!h]
\includegraphics[scale=.75]{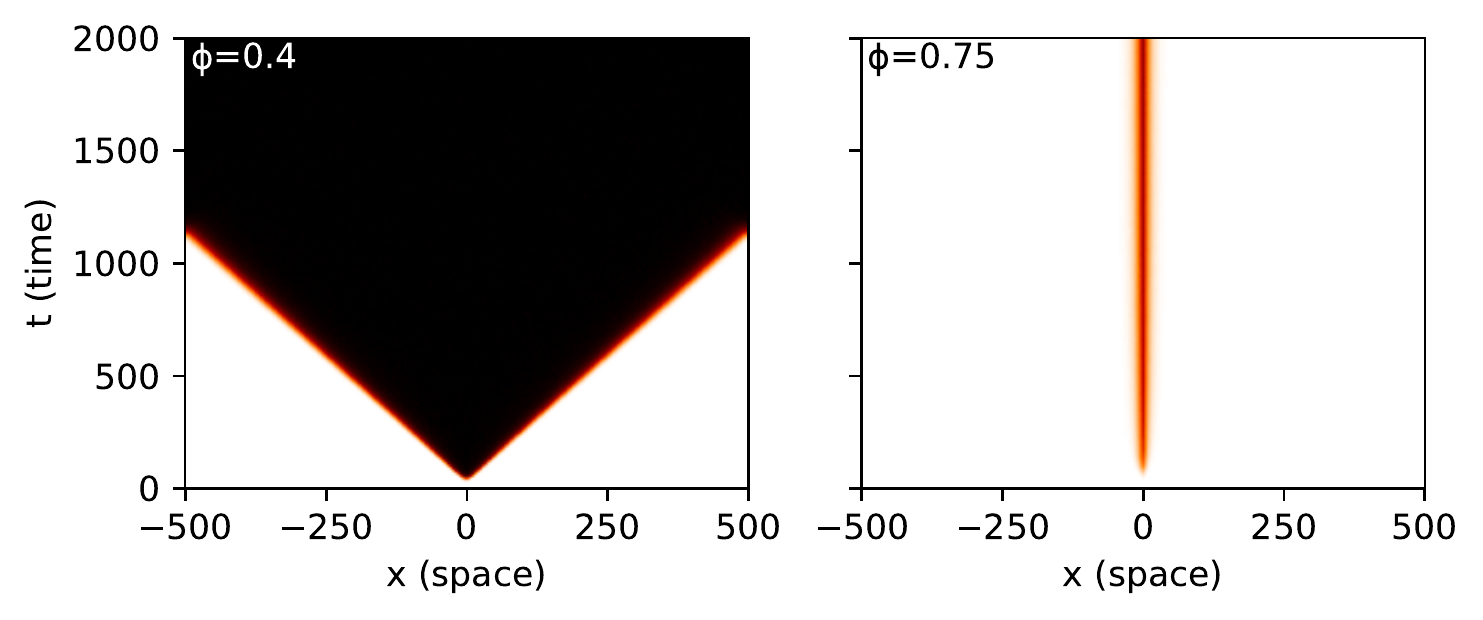}
\caption{Results for the classical OTOC for the time-independent Hamiltonian in Eq.~\ref{eq:Hising} as heatmap in spacetime; white denotes 0 and black denotes 1. The left panel corresponds to the chaotic phase ($\phi<\phi_c$) whereas the right panel corresponds to the localised phase ($\phi>\phi_c$).}
\label{fig:hamresults}
\end{figure}

We propagate these equations of motion using RK4 with a timestep of 0.1 and the classical OTOC is computed in exactly the same fashion as in the main text. The results are shown in Fig.~\ref{fig:hamresults} where it is clear that for large enough values of $\phi$, the OTOC is completely arrested such that there again exists a bonafide dynamical phase transition.

\section{Results in two spatial dimensions \label{sec:2d}}
\begin{figure}
\includegraphics[scale=.75]{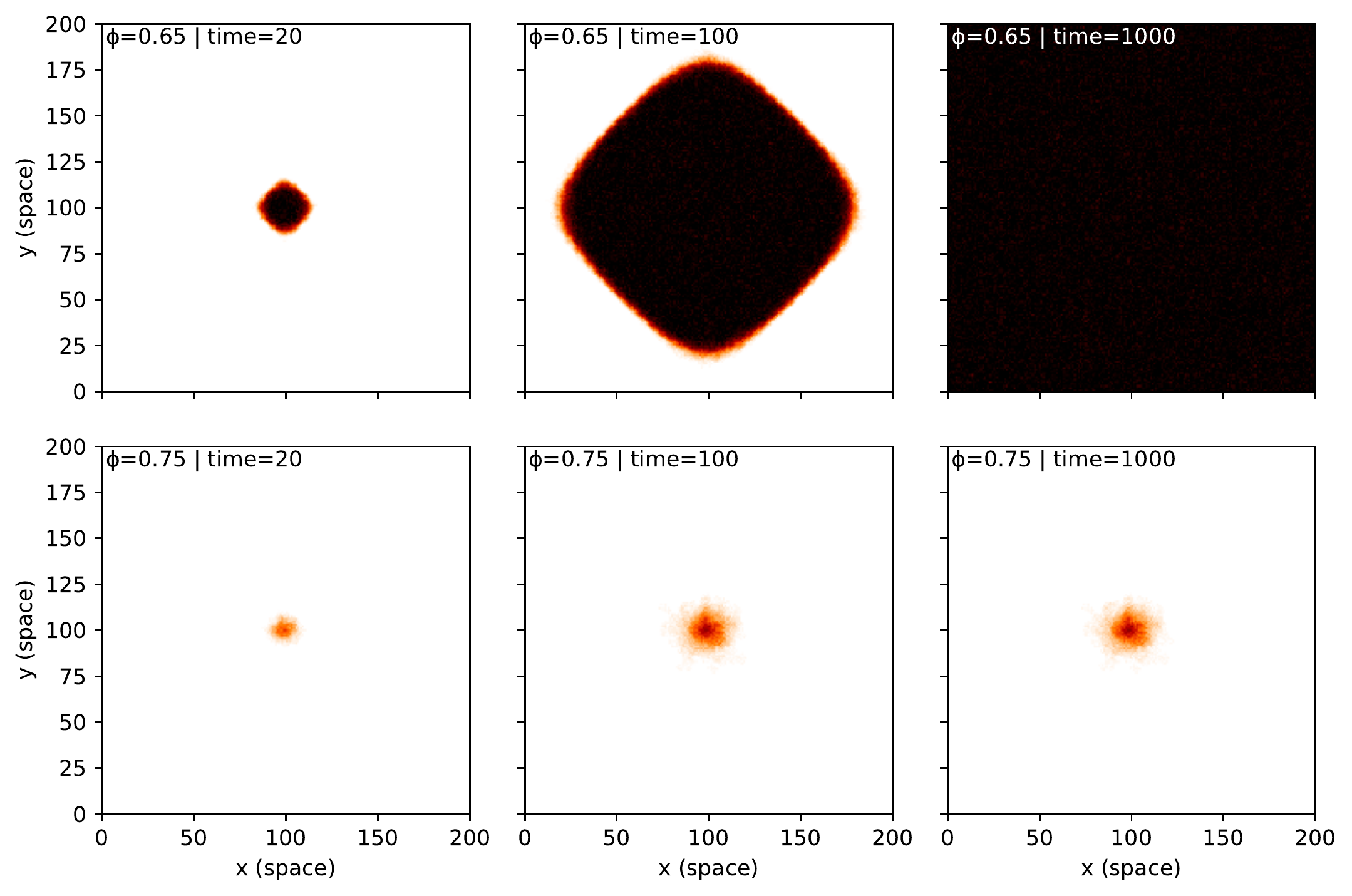}
\caption{Results for the classical OTOC for the two-dimensional model described in Eq.~\ref{eq:hamiltonian2d} as heatmaps in two-dimensional space. The two rows correspond to two values of $\phi$ and the different columns correspond to different times increasing from left to right. The top row corresponds to $\phi<\phi_c$ and the OTOC spreads ballistically whereas the bottom row corresponds to $\phi>\phi_c$ the OTOC stops spreading after a finite time.}
\label{fig:2dresults}
\end{figure}

In this section, we present results for a Floquet model, analogous to that in Eq.~1 in the main text, but defined on a square lattice in two spatial dimensions. The time-periodic Hamiltonian is described by 
\begin{equation}
  \mathcal{H}(t)=
  \begin{cases}
  \sum\limits_{x,y}\sum\limits_{\eta=\pm1} [\s{z}{x,y}(\s{z}{x+\eta,y}+\s{z}{x,y+\eta}+h )],~t \in[nT, (n+\frac{1}{2})T)\\
  \sum\limits_{x,y}  g \s{x}{x,y}\,,~~t\in [(n+\frac{1}{2})T,(n+1)T)
  \end{cases}\,.
\label{eq:hamiltonian2d}
\end{equation}
The stroboscopic equations of motions have exactly the form as in Eq.~2 of the main text,
\begin{equation}
\mathbf{S}_{x,y}[(n+1)T] = \begin{cases} \mathsf{R_x}[gT/2]\mathsf{R_z}[\theta_{x,y}(nT)] \mathbf{S}_{x,y}(nT);~\Theta_{x,y}(nT)=1\\
\mathbf{S}_{x,y}(nT);~\Theta_{x,y}(nT)=0
\end{cases}\,,
\end{equation}
except the rotation angle $\theta_{x,y} = \sum_{\eta=\pm1}[S^z_{x+\eta,y}+S^z_{x,y+\eta}+h]$ now depends on all the four neighbours of the spin at site $(x,y)$. Similarly, the constraint also depends on the configurations of all the four neighbours,
\begin{equation}
\Theta_{x,y}(nT)=\Theta[\cos(\pi\phi)-\min(S_{x-1,y}^z(nT),S_{x+1,y}^z(nT),S_{x,y-1}^z(nT),S_{x,y+1}^z(nT))]\,.
\label{eq:constraints2d}
\end{equation}
Physically, it implies that for a spin at any site to be active, it is sufficient for one of its four neighbours to be outside the constraining cone. In Fig.~\ref{fig:2dresults}, we present exemplary results for two values of $\phi$, one in the chaotic phase and one in the arrested phase, again cofirming that the transition is present also in two spatial dimensions.

\end{document}